\documentclass[reprint,amsmath,amssymb,aps,superscriptaddress]{revtex4-2}

\usepackage{graphicx}
\usepackage{dcolumn}
\usepackage{bm}

\begin{document}

\title{Optically induced anisotropy in time-resolved scattering: Imaging molecular scale structure and dynamics in disordered media with experiment and theory}

\author{Andr\'es Montoya-Castillo}
\affiliation{Department of Chemistry, University of Colorado, Boulder, Boulder, Colorado, 80309, USA}

\author{Michael S. Chen}
\affiliation{Department of Chemistry, Stanford University, Stanford, California, 94305, USA}

\author{Sumana L. Raj}
\affiliation{Stanford PULSE Institute, SLAC National Accelerator Laboratory, Stanford University, Menlo Park, California 94025, USA}

\author{Kenneth A. Jung}
\affiliation{Department of Chemistry, Stanford University, Stanford, California, 94305, USA}

\author{Kasper S. Kjaer}
\affiliation{Stanford PULSE Institute, SLAC National Accelerator Laboratory, Stanford University, Menlo Park, California 94025, USA}

\author{Tobias Morawietz}
\affiliation{Department of Chemistry, Stanford University, Stanford, California, 94305, USA}

\author{Kelly J. Gaffney}
\email{kgaffney@slac.stanford.edu}
\affiliation{Stanford PULSE Institute, SLAC National Accelerator Laboratory, Stanford University, Menlo Park, California 94025, USA}

\author{Tim B. van Driel}
\email{timbvd@slac.stanford.edu}
\affiliation{Linac Coherent Light Source, SLAC National Accelerator Laboratory, Menlo Park, California 94025, USA}

\author{Thomas E. Markland}
\email{tmarkland@stanford.edu}
\affiliation{Department of Chemistry, Stanford University, Stanford, California, 94305, USA}

\date{\today}

\begin{abstract}
Time-resolved scattering experiments enable imaging of materials at the molecular scale with femtosecond time resolution. However, in disordered media they provide access to just one radial dimension thus limiting the study of orientational structure and dynamics. Here we introduce a rigorous and practical theoretical framework for predicting and interpreting experiments combining optically induced anisotropy and time-resolved scattering. Using impulsive nuclear Raman and ultrafast X-ray scattering experiments of chloroform and simulations, we demonstrate that this framework can accurately predict and elucidate both the spatial and temporal features of these experiments. 
\end{abstract}

\maketitle

The characterization of the structure and dynamics of disordered media ranging from liquids to solutions and glasses with atomic resolution remains a formidable challenge. Whether using X-rays, neutrons, or electrons to characterize the structure of these systems the orientational averaging associated with isotropic samples reduces the structural information to a single radial dimension inhibiting the ability to resolve the orientational structure and dynamics.

The situation becomes even more vexing when the goal extends to understanding the dynamics of liquids and solutions with atomic resolution. Accessing this information requires techniques that can probe timescales from tens of femtoseconds to hundreds of picoseconds corresponding to frequencies ranging from $\sim$0.1 to $\sim$1,000 cm$^{-1}$. While ultrafast nonlinear spectroscopies can be used to investigate dynamics in this temporal range, they do not provide a direct link to the underlying structural changes occurring, necessitating their combination with other techniques, such as simulation, to infer these connections \cite{Mukamel1995}. In contrast, inelastic X-ray and neutron scattering measurements provide access to the required spatial resolution, but in practice the low count rates of the measurements limit their impact~\cite{Sette1998,Teixeira1985,Teixeira1985a}.

The advent of intense ultrafast X-ray laser sources has enabled the combination of ultrafast non-linear optical excitation methods and X-ray scattering to access the dynamics of liquids and solutions in the femtosecond temporal and {\AA}ngstr{\"o}m spatial dimensions~\cite{Kim2021,Chergui2017,Gaffney2021,Biasin2021}. In particular, impulsive nuclear Raman and X-ray scattering (INXS) combines non-resonant impulsive stimulated Raman scattering (ISRS) interactions, which excite Raman active vibrational and rotational motions~\cite{Vohringer1995,Dhar1994,Loughnane1999} and generate non-equilibrium nuclear geometries, with ultrafast time-resolved elastic X-ray scattering to probe these structural dynamics~\cite{Ki2021,Kim2020}. INXS has recently been experimentally realized ~\cite{Biasin2016,Ki2021,Kim2020} and offers the potential to obtain radial and orientational structure and dynamics of liquids on the {\AA}ngstr{\"o}m to nanometer length scale and the femtosecond to picosecond time scale. By generating changes in both the radial and orientational order of disordered media, ISRS interactions can generate both isotropic and anisotropic changes in the structure factor, thus extending radial information present in INXS to a second spatial (orientational) dimension. These developments thus present a tremendous opportunity for combining optically induced anisotropic excitations with nuclear scattering probes such as ISRS with X-ray scattering to transform our understanding of the structure and dynamics of disordered media.

To elucidate and harness the rich information present in experiments which exploit the combination of optically induced anisotropy and time-resolved scattering and inspire future developments, it is essential to develop quantitatively predictive theories that can reproduce experimental signals and provide an interpretational framework of these cutting edge measurements. Here we derive a rigorous and practical quantum mechanical framework in the linear response limit for predicting and interpreting experiments involving optically induced anisotropy combined with time-resolved scattering. By performing INXS experiments of liquid cholorform and simulations of our theory based on a polarizable force field~\cite{Mu2014}, we show that our approach provides a quantitative description of the INXS signal and thus provides a robust and clear foundation for understanding and interpreting INXS experiments that can be extended to other liquids and solutions.

From both a theoretical and interpretational perspective, the linear response formulation \cite{Mukamel1995} offers numerous advantages. By employing this formulation we cast the system's nonequilibrium response to light-matter interactions in terms of equilibrium time correlation functions, which allows one to elucidate a system's inherent response rather than one that depends on the details of the nonequilibrium simulation designed to mirror the physical process being probed. This formulation incorporates the optical and X-ray pulses through a convolution with the system's inherent response, allowing one to also study the effect of pulse shaping on the signal. When taken to the classical limit of the nuclei, our approach permits the use of the well developed toolbox of atomistic simulation techniques to simulate the INXS signal.

To formulate our theory, we consider the features of the INXS experiment. INXS initially pumps the sample with an electronically and vibrationally off-resonant optical pulse and then probes it with an off-resonant X-ray pulse, leading to a time-resolved scattering pattern on the detector plane (SI~Fig.1). The frequency dependence of the INXS signal recovers Raman active modes (see Fig.~\ref{fig:s2_q}). Hence, INXS is a third-order nonlinear scattering method with the initial interaction consisting of an optical Raman excitation followed by a scattering interaction~\cite{Ki2021,Kim2020}. 

We thus employ an expansion of the system's density matrix to third-order in the light-matter coupling~\cite{Mukamel1995,Tanaka2001} subject to the INXS sequence of excitation and scattering interactions. In SI Sec.~1, we first derive a rigorous quantum mechanical expression for the INXS signal and then demonstrate how INXS experiments permit a series of well-controlled approximations \cite{Cho1993,Yan1991,Tanimura1993,Luber2014,Long2002, Dixit2012,Moller2012,Helliwell1997, Coppens1992,Ben-Nun1997,Cao1998,Rozgonyi2005,Dohn2015, MorenoCarrascosa2017,Northey2014,Northey2016,MorenoCarrascosa2019, Mukamel1995,Egorov1999,Craig2004, Craig2005,Ramirez2004} that result in an expression for the INXS signal at time $t$ after the initial excitation that is compatible with direct atomistic simulations,
\begin{equation}\label{eq:inxs-signal}
\begin{split}
    \Delta S(\mathbf{q}_{xy}, t) &= \mathcal{N}^{\rm clas} \int_0^{t} d\tau\ |E_{\rm opt}(t - \tau)|^2  \mathcal{R}(\mathbf{q}_{xy}, \tau),
\end{split}
\end{equation}
where $\mathcal{N}^{\rm clas}$ is given in Eq.~\ref{eq:proportionality-constant}, $E_{\rm opt}(t)$ is the electric field envelope for the optical interaction and 
\begin{equation}\label{eq:inxs-response}
    \mathcal{R}(\mathbf{q}_{xy}, t) = \frac{d}{dt} \langle \tilde{\alpha}(0, \{ \mathbf{R}\}) \Sigma(\mathbf{q}_{xy}, t) \rangle_{eq}.
\end{equation}
Here $\mathcal{R}(\mathbf{q}_{xy}, t)$ is the INXS response function in the limit of classical nuclei (for the quantum mechanical expression, see SI Eq.~37) that correlates the Raman excitation at $t=0$, given by the linear combination of polarizability tensor elements, with the X-ray scattering pattern at time $t$, given by the $x-y$-cut (perpendicular to the $z$-propagation-direction of the incoming X-ray probe) of the X-ray scattering operator, $\Sigma(\mathbf{q}_{xy})$. While our expression can be combined with a range of methods to treat $\Sigma(\mathbf{q}_{xy})$, in our simulations below we employ the widely adopted independent atom model \cite{Moller2012,Helliwell1997}. 

An immediate success of our expression for the INXS signal, Eq.~\ref{eq:inxs-signal}, lies in the derived proportionality constant, 
\begin{equation} \label{eq:proportionality-constant}
    \mathcal{N}^{\rm clas} = -\frac{\beta (\mathbf{e}_{\rm xr} \cdot \mathbf{e}_{\rm s})^2 |E_{\rm xr}|^2 |E_{\rm s}|^2}{\omega_{\rm opt}^2 \omega_{\rm s}^2 \omega_{\rm xr}^2 }
\end{equation} 
which offers testable predictions of the scaling of the INXS signal with respect to the experimental setup and strengths and frequencies of the light pulses. In particular, $\mathcal{N}^{\rm clas}$ predicts that the INXS signal is strongest when the incoming and scattered X-ray are collinearly polarized, ($\mathbf{e}_{\rm xr} \cdot \mathbf{e}_{\rm s} = 1$), and diminishes to zero when perpendicular ($\mathbf{e}_{\rm xr} \cdot \mathbf{e}_{\rm s} = 0$). It also depends quadratically on the magnitude of the electric field for the optical ($E_{\rm opt}$) and incoming ($E_{\rm xr})$ and scattered ($E_{\rm s}$) X-ray pulses, and is inversely proportional to the square of the frequencies for the optical ($\omega_{\rm opt}$) and incoming ($\omega_{\rm xr}$) and scattered ($\omega_{\rm s}$) X-ray pulses. Our experiments have confirmed all these predictions, providing support for applicability of our linear response theory treatment of the INXS interactions.

A fundamental question that arises in the theoretical treatment of the INXS experiment is when an anisotropic INXS signal will emerge from a homogeneous and isotropic system. Equation~\ref{eq:inxs-response} shows that the anisotropic signal arises from the instantaneous correlation of the components of the polarizability tensor, $\alpha_{\boldsymbol{\epsilon}_{\rm opt}}(\{ \mathbf{R}\})$, that depend on the global nuclear configuration, $\{\mathbf{R}\}$, and the $x-y$ cut of the modified structure factor, $\mathbf{q}_{xy} = q_x\hat{\mathbf{x}} + q_y\hat{\mathbf{y}}$, that arises from a constrained Fourier transform ($q_z = 0$) of the microscopic positions of all atoms in the system. Equation~\ref{eq:inxs-response} also shows that systems characterized by an isotropic polarizability tensor, $\alpha_{xx} = \alpha_{yy} = \alpha_{zz}$ and $\alpha_{jk} = 0$ where $j \neq k \in \{x, y, z \}$, can only produce isotropic INXS signals. 

This is because the linear combination of the polarizability tensor elements in Eq.~\ref{eq:inxs-response} is set by the polarization of the optical pulse, $\boldsymbol{\epsilon}_{\rm opt} = \hat{\mathbf{x}} \cos(\phi)  + \hat{\mathbf{y}}\sin(\phi) $, which is taken to be in the $x-y$ plane (see SI Fig.~1), and leads to $\tilde{\alpha} = \cos^2(\phi) \alpha_{xx} + \sin^2(\phi) \alpha_{yy} + \sin(\phi)\cos(\phi)(\alpha_{xy} + \alpha_{yx})$. This is because the polarization of the optical pulse, whether it is linearly or circularly polarized, sets the linear combination of the polarizability tensor elements in Eq.~\ref{eq:inxs-response} (SI Eq.~33). For instance, an optical pulse linearly polarized at an angle $\phi$ with respect to the $x$-$y$ plane, $\boldsymbol{\epsilon}_{\rm opt} = \hat{\mathbf{x}} \cos(\phi)  + \hat{\mathbf{y}}\sin(\phi) $, as is the case in our experiment (see SI Fig.~1),  leads to $\tilde{\alpha} = \cos^2(\phi) \alpha_{xx} + \sin^2(\phi) \alpha_{yy} + \sin(\phi)\cos(\phi)(\alpha_{xy} + \alpha_{yx})$. As such, when $\alpha_{xx} = \alpha_{yy}$ and $\alpha_{xy} = 0 = \alpha_{yx}$, the resulting INXS signal would have no $\phi$ dependence on the detector plane and therefore no anisotropic component. Hence, the anisotropy of the INXS signal arises from the filtering of the wavevector-restricted structure factor, $\Sigma(\mathbf{q}_{xy})$, imposed by the anisotropy of the polarizability tensor, $\alpha_{\boldsymbol{\epsilon}_{\rm opt}}(\{ \mathbf{R}\})$, of the entire system which depends on the global nuclear configuration of the sample. As the experimental and simulation results below demonstrate, the INXS signal provides an avenue to break the symmetry of an X-ray scattering pattern as filtered through the nuclear position-dependent polarizability tensor and analyze how its time evolution reveals the connection between structural rearrangements and energy relaxation pathways in complex liquids. 

\begin{figure}[]
    \begin{center}
        \includegraphics[width=0.48\textwidth]{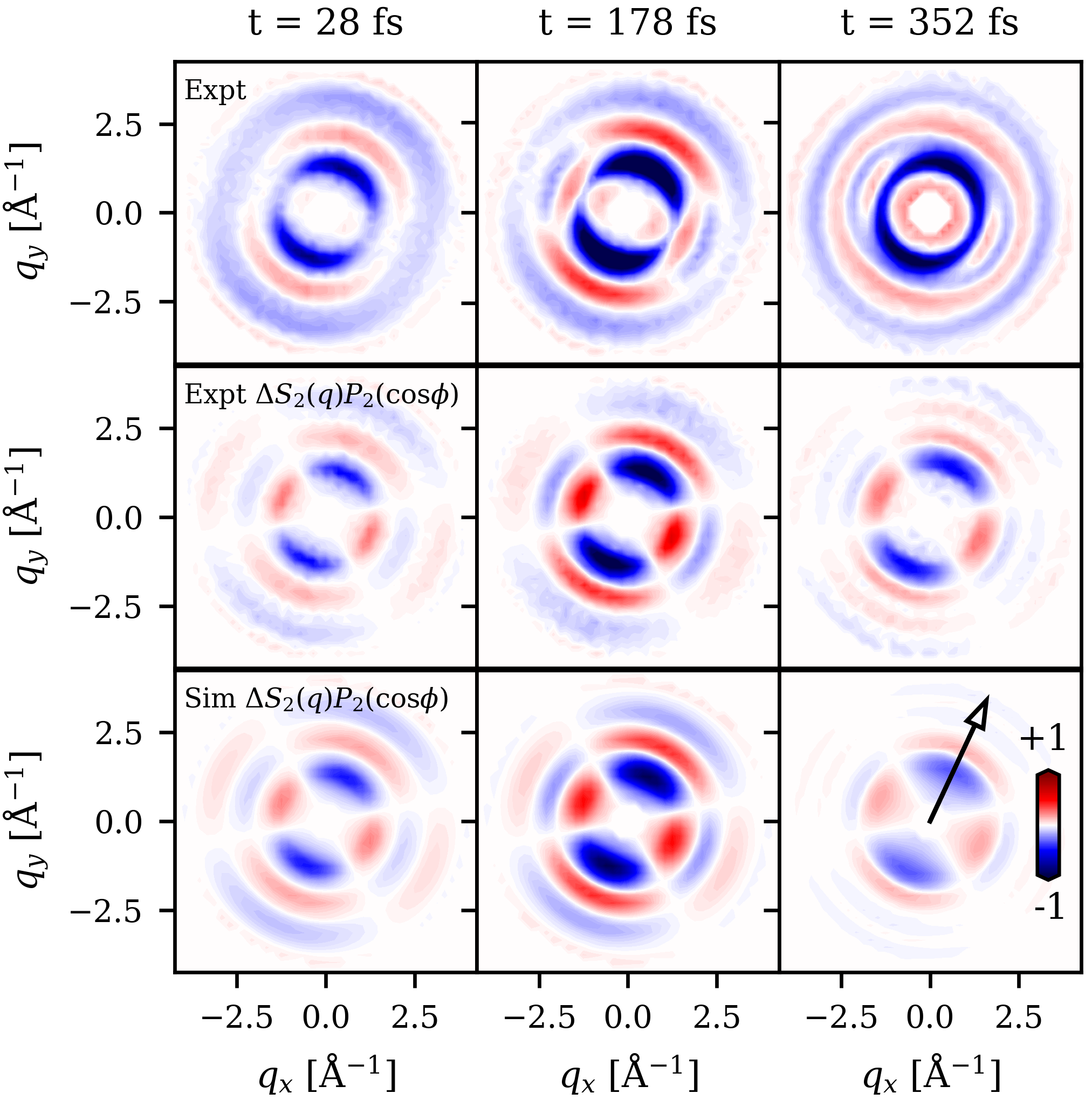}
    \end{center}  
    \vspace{-6mm}
    \caption{The difference experimental INXS signal (top, see SI Fig.~6 for raw signal) and the anisotropic component, $\Delta S_2(q,t)P_{2}(\cos{\phi})$, of the signal obtained from experiment (middle) and theory (bottom) at times from 28~fs to 352~fs following the initial excitation of the sample. The arrow in the lower right panel defines the direction of the optical excitation electronic field. q$_{y}$ and q$_{x}$ are the axes of the gridded two-dimensional difference scattering signal. Red denotes an increase and blue denotes a decrease in the signal relative to the equilibrium scattering.}
	\label{fig:full_inxs}
	\vspace{-4mm}
\end{figure}

To assess the ability to understand and interpret INXS experiments using this theoretical approach we performed INXS experiments (SI Sec.~2) at the XPP endstation of the LCLS X-ray free electron laser (XFEL). The pump was an 800 nm, 50 fs full-width at half-maximum (FWHM) ultrafast laser pulse and the probe was an 8.8 keV monochromatic, 40 fs FWHM X-ray pulse. Both pulses were linearly polarized with about a 30\textdegree angle between the two polarizations. The sample, liquid chloroform (CHCl$_3$) at room temperature, was delivered to the interaction point with a 50 \textmu m diameter fused silica capillary based cylindrical jet. We detected the scattered X-rays  with the CSPAD large area detector~\cite{Hart2012}, for scattering vectors ($q$) up to 4.5 \AA$^{-1}$. The experimental time resolution was measured to be $\sim$70 fs FWHM using a water sample as reference. During the experiment, it was verified that the anisotropic difference scattering signal increases linearly with pump pulse energy.

The top row of Fig.~\ref{fig:full_inxs} compares the difference INXS signal, $\Delta S(q,t)$, relative to the equilibrium elastic X-ray scattering signal, measured at times from 28~fs to 352~fs after excitation. This signal exhibits strong anisotropy that increases for  the first 200~fs, followed by a decay over the next few picoseconds that leaves an isotropic background signal due to an increase in liquid temperature of $\sim$18~K arising from the pump excitation (SI Sec.~2 and SI Fig.~2). The prominent anisotropic nature of the signal highlights the utility of this approach in obtaining orientationally resolved structure and dynamics in the isotropic sample. To isolate the anisotropic component, we exploit the fact that the INXS signal can be accurately decomposed~\cite{Natan2021,Biasin2018,Lorenz2010,Baskin2006,Baskin2005,VanKleef1999,Ben-Nun1997} in terms of just the zeroth (isotropic) and second order (anisotropic) Legendre polynomials with negligible higher order contributions (SI Fig.~3 and 4). The ability to decompose the signal into the zeroth and second order Lengendre polynomials alone is consistent with what is expected for ISRS and the theory formulated in SI Sec.~1. The middle and bottom rows of Fig.~\ref{fig:full_inxs} show the anisotropic contribution to the INXS signal ($\Delta S_2$) obtained from both experiment and molecular dynamics simulations (see SI Sec.~4) of the theoretical treatment of INXS introduced here. The excellent agreement between the experimental and simulated $\Delta S_2$ across the full time and scattering wavevector ($q$) ranges probed demonstrates that the anisotropic component of the INXS signal can be accurately captured within linear response using our theoretical approach when combined with a polarizable force field treatment of chloroform.

To provide further insight into the time evolution of the anisotropic ($\Delta S_2$) component of the INXS difference signal, Fig.~\ref{fig:s2_q} shows the scattering wavevector ($q$) dependence of the signal for 1.2~ps following the initial excitation. Here the excellent agreement between the experiment and theory across the entire time range shown is even more evident with both exhibiting pronounced oscillatory features in the high $q$ region while the low $q$ region exhibits more non-oscillatory features indicative of diffusive dynamics. To elucidate the physical origins of these oscillatory features, the right hand panel of Fig.~\ref{fig:s2_q} contains the frequency dependence of the anisotropic signal (see SI Sec.~6). From this we observe that the anisotropic signal has a large component at 261~cm$^{-1}$ for the experiment and at 234~cm$^{-1}$ for the simulation. The source of this frequency can be explained by considering the normal modes of the chloroform molecule (SI Table.~1). Of chloroform's nine normal modes, six fall outside the current resolvable frequency range of the experiment ($<400$ cm$^{-1}$), due to the temporal width of the excitation pulse. Of the three remaining normal modes, one is a doubly degenerate ($e$ symmetry) vibration and appears at a frequency of 261~cm$^{-1}$  with the other an ($a_1$ symmetry) vibration at 366~cm$^{-1}$ in the experimental Raman spectrum of the liquid~\cite{Madigan1951}. We note that in the force field simulation, these modes appear in the Raman spectrum at 234~cm$^{-1}$ and 298~cm$^{-1}$ respectively (SI Fig.~5). The frequencies and symmetries of the intramolecular vibrations observed with INXS match the expectations established by polarized Raman scattering and ISRS. Specifically, the lower frequency asymmetric CCl bend (261 cm$^{-1}$, $e$ symmetry) should predominantly appear in the anisotropic INXS signal, while the higher frequency symmetric CCl bend (366 cm$^{-1}$, $a_1$ symmetry) should predominantly appear in the isotropic INXS signal~\cite{Madigan1951}. The experimental and simulated INXS signals confirm these expectations (right panels of Fig~\ref{fig:s2_q}). However, to understand the less oscillatory anharmonic intermolecular modes probed by the experiment it is useful to consider how the signal varies in time as a function of position. 

\begin{figure}[]
    \begin{center}
        \includegraphics[width=0.48\textwidth]{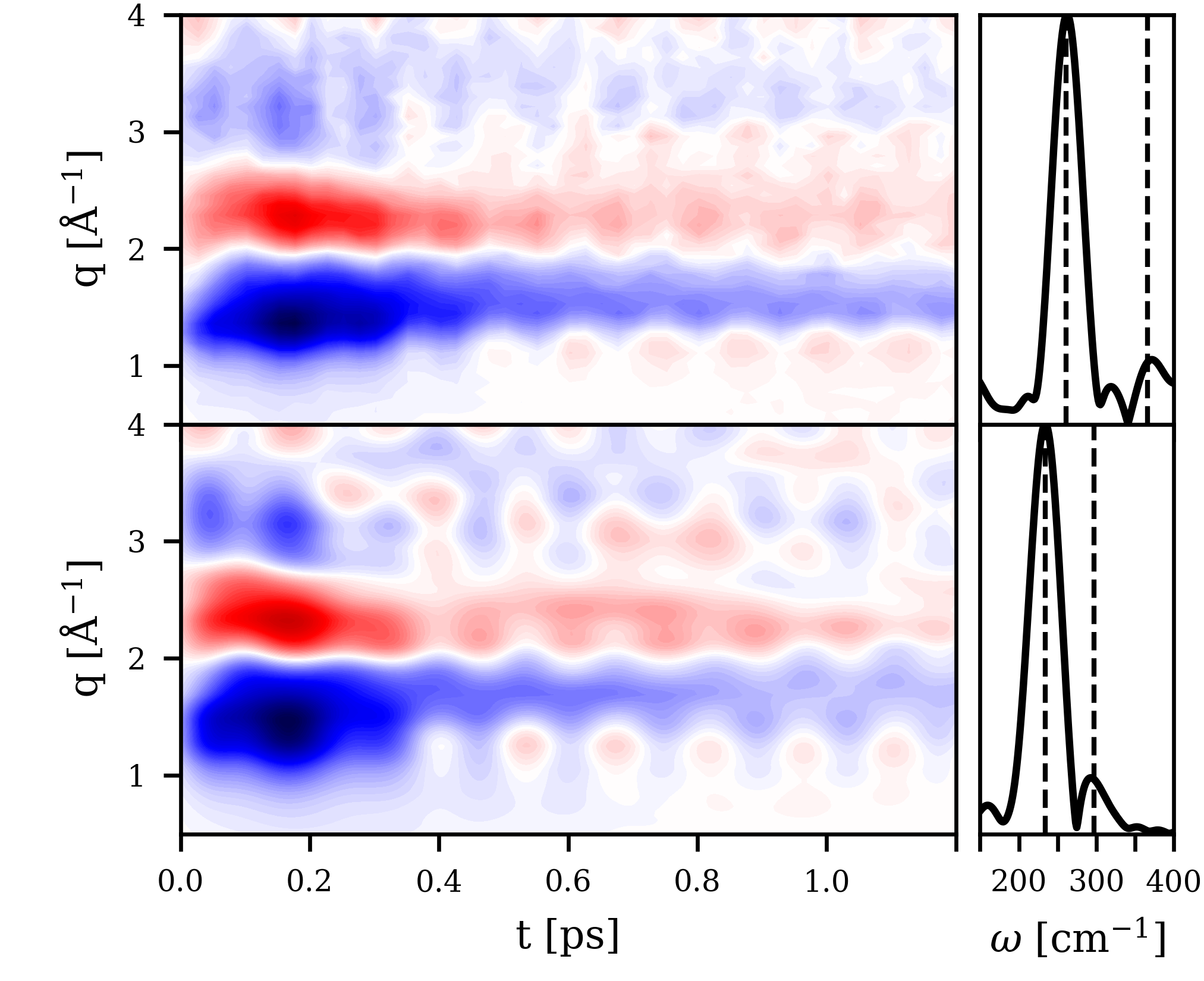}
    \end{center}  
    \vspace{-6mm}
    \caption{The anisotropic component of the INXS signal, $\Delta S_2(q,t)$, from both the experiment (top left) and the simulation (bottom left). Their respective Fourier transforms (right) show the oscillations in time of the anisotropic component are primarily due to the lowest frequency vibrational mode for chloroform which occurs at 261~cm$^{-1}$ for the experiment and at 234~cm$^{-1}$ for the simulation. These frequencies are marked as the leftmost dashed line in their respective plots. The second lowest frequency mode is also marked and only weakly contributes to the anisotropic signal.}
	\label{fig:s2_q}
    \vspace{-5mm}
\end{figure}

\begin{figure*}[]
    \begin{center}
        \includegraphics[width=\textwidth]{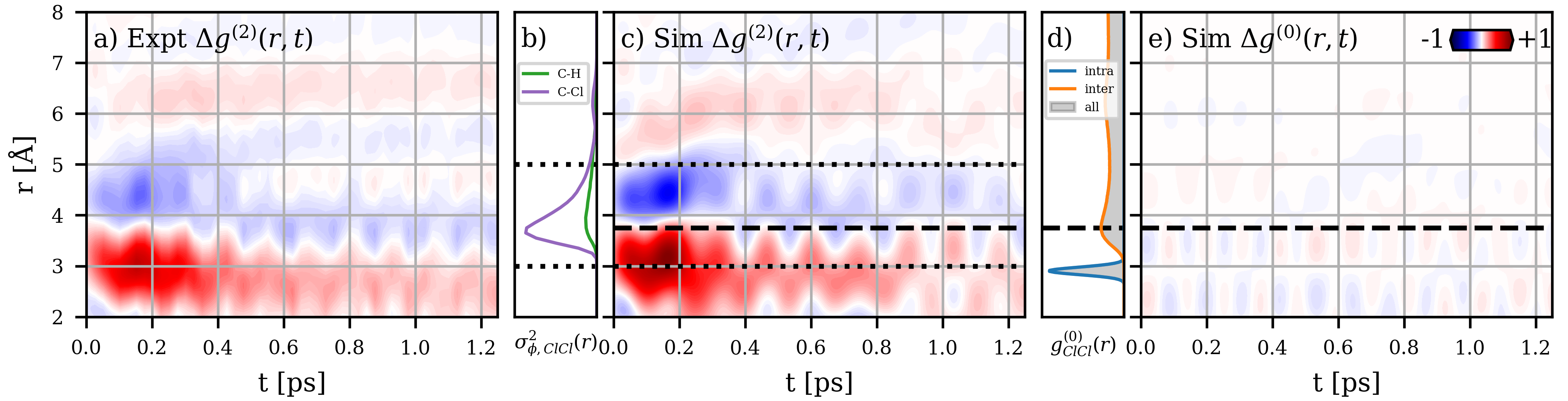}
    \end{center}
    \vspace{-8mm}
    \caption{The real space transformed anisotropic component of the INXS signal, $\Delta g^{(2)}(r,t)$, from (a) experiment and (c) simulation. (b) Simulated variance arising from angular fluctuations at each distance along the C-H and C-Cl axes. Comparing $\Delta g^{(2)}(r,t)$ in (c) with the simulated chlorine-chlorine radial distribution function $g_{ClCl}^{(0)}(r)$ in (d) demonstrates that the anisotropic non-oscillatory component is intermolecular in character since it straddles the first intermolecular (orange) peak in $g^{(0)}(r)$ but is absent from the isotropic component $\Delta g^{(0)}(r,t)$ shown in (e).
    }
	\label{fig:g2_g0}
    \vspace{-5mm}
\end{figure*}

To discern the more diffusive intermolecular contributions from the more oscillatory intramolecular contributions to the INXS signal, Fig.~\ref{fig:g2_g0} shows results of transforming both the isotropic and anisotropic components to real space where the spatial extent of these features helps identify their origins. Since chlorine atoms are by far the largest scatterers in this system, with an atomic form factor nearly three-fold greater than that of the other atoms, one would expect the INXS signal to be dominated by chlorine-chlorine scattering. This is indeed the case in the simulations where, as demonstrated in SI Fig.~7, we can exactly extract the chlorine-chlorine contributions to the signal to leave only a small residual. By exploiting this realization, as shown in SI Sec.~7, the transformation of the experimental signal to real space and its physical interpretation can be simplified considerably.

While our primary focus is the physical interpretation of the anisotropic information that INXS reveals, it is instructive to first consider the content of the isotropic component. In particular, one can connect the isotropic component, $\Delta S_{0}(q,t)$, to the difference pair distribution function
\begin{subequations}
\begin{align}
    \Delta g^{(0)}_{ClCl}(r,t) &= 2\int_{0}^{\infty}\text{d}q~q^{2}j_{0}(qr) \frac{\Delta S_{0}(q, t)}{F_{Cl}^{*}(q)F_{Cl}(q)} \label{eq:g0_rt} \\
    &= \int_{0}^{\pi} \text{d}\varphi~\sin{\varphi}\Delta g_{ClCl}(r,\varphi, t),\label{eq:g0_rt-Legendre}
\end{align}
\end{subequations}
where $j_{n}$ is the $n$-th order spherical Bessel function (see SI Sec.~7). Equation~\ref{eq:g0_rt-Legendre} introduces the well known expression for the angular average over the polar angle $\varphi$ (see SI Sec.~7), which in the case of a conventional X-ray scattering signal corresponds to the circularly symmetric signal on the detector plane. Applying Eq.~\ref{eq:g0_rt} to the isotropic component of the simulated signal yields the difference pair distribution function shown in Fig.~\ref{fig:g2_g0}e, where we see oscillations from the 293~cm$^{-1}$ chloroform mode centered around the first intramolecular peak of $g_{ClCl}(r)$.

A similar transformation gives access to the real space information contained in the anisotropic INXS signal. Specifically, we apply the second order Bessel transformation to $\Delta S_2(q,t)$, 
\begin{subequations}
\begin{align}
    \Delta g^{(2)}_{ClCl}(r,t) &= -2\int_{0}^{\infty}\text{d}q~q^{2}j_{2}(qr) \frac{\Delta S_{2}(q, t)}{F_{Cl}^{*}(q)F_{Cl}(q)}\\
    &= \int_{0}^{\pi} \text{d}\varphi~\sin{\varphi}\Delta g_{ClCl}(r,\varphi, t) P_{2}(\cos{\varphi}),\label{eq:g2_rt-Legendre}
\end{align}
\end{subequations}
By comparing the expression in Eq.~\ref{eq:g2_rt-Legendre} to the expression for the standard difference pair distribution function in Eq.~\ref{eq:g0_rt-Legendre} and noting that $P_0(\cos(\varphi)) = 1$, we see that that this procedure yields a difference generalized pair distribution function that encapsulates the angularly resolved fluctuations of the distribution following the symmetry of the second-order Legendre polynomial $P_2(\cos(\varphi))$. $\Delta g^{(2)}_{ClCl}(r,t)$ thus provides a direct measure of fluctuations in the projection of the pair distribution onto the anisotropic polarizability. This anisotropic component of the INXS signal mirrors the symmetry of the second-order Legendre polynomial, which displays a pattern with two pairs of positive and negative lobes around the circle, and when fully averaged across the circle would disappear. This additional information expands on that provided by conventional X-ray and neutron scattering experiments which only provide the isotropic averaging shown in Eq.~\ref{eq:g0_rt-Legendre}. The transformation of the simulated $\Delta S_2(q,t)$ to real space, $\Delta g^{(2)}(r,t)$, as shown in Fig.~\ref{fig:g2_g0}c allows us to determine the origins of the non-oscillatory features in the signal. For example, the non-oscillatory feature that peaks at $\sim$200~fs and subsequently decays straddles the first intermolecular chlorine-chlorine peak of the simulated $g_{ClCl}(r)$ centered at 3.75~$\text{\AA}$. This suggests that it arises from motions following excitation that lead to purely intermolecular redistribution. This is confirmed in SI Fig.~8 where we exactly decompose the simulated INXS signal into its intra- and intermolecular components where the intramolecular component lacks this feature. In contrast, this non-oscillatory intermolecular peak does not appear in the isotropic signal. Hence, the Raman excitation causes an anisotropic orientational rearrangement of neighboring chloroform molecules (angularly distributed according to the second-order Legendre polynomial) that conserves the average intermolecular chlorine-chlorine distance.  
To provide insight into the local molecular arrangements that lead to the non-oscillatory positive and negative feature at $\sim$200~fs in the region 3-5~$\text{\AA}$ in $\Delta g^{(2)}(r,t)$, we consider the density fluctuations of chlorine that dominate the anisotropic response in the real space INXS signal. Since $g^{(2)}(r,t)$ arises from anisotropic orientational rearrangements, the experimental signal at a particular distance $r$ is sensitive to the variance of the angular distribution (Fig.~\ref{fig:g2_g0}b) at that distance, i.e. if there is no angular variation in the molecular density at that distance, the signal will be isotropic and will not contribute to $\Delta g^{(2)}(r,t)$. Figure~\ref{fig:g2_g0}b shows the variance of the angular distribution of the chlorine-chlorine radial distribution function along the two principal symmetry axes of the chloroform molecule: the C-H axis and the Cl-C-Cl bisector perpendicular to the C-H axis (see SI Sec.~10). From this, one can see that for both axes the variance of the angular distribution is largest in the region 3-5~$\text{\AA}$ consistent with the regions of the greatest signal in $g^{(2)}(r,t)$ (dotted lines in Fig.~\ref{fig:g2_g0}c). This can be seen even more clearly in SI Fig.~7 where the we show just the intermolecular $\Delta g^{(2)}(r,t)$ obtained from the simulations. 
Thus, by breaking the symmetry of an isotropic sample through the directionality of a Raman interaction, the real space anisotropic INXS signal reveals an angularly resolved microscopic density and orientational fluctuations in chemical, biological, and materials systems with fs and \AA~level resolution as a response to an impulsive excitation. The theoretical and simulation framework presented here offers the opportunity to obtain and understand the unique orientational structure and dynamical information about the molecular to nanoscopic structure of liquids and glasses present in experiments that harness optically induced anisotropy in time-resolved scattering.

\vspace{2mm}
This work was supported by the U.S. Department of Energy, Office of Science, Basic Energy Sciences, Chemical Sciences, Geosciences, and Biosciences Division. The authors wish to acknowledge LCLS staff including T. Sato, J. Glownia, D. Zhu, S. Nelson and A. Robert for experimental support and helpful discussion. Use of the Linac Coherent Light Source (LCLS), SLAC National Accelerator Laboratory, is supported by the U.S. Department of Energy, Office of Science, Office of Basic Energy Sciences under Contract No. DE-AC02-76SF00515. This research used resources of the National Energy Research Scientific Computing Center (NERSC), a U.S. Department of Energy Office of Science User Facility operated under Contract No. DE-AC02-05CH11231.

\bibliography{bibliography}

\end{document}


\title{Supplementary material for: Optically induced anisotropy in time-resolved scattering: Imaging molecular scale structure and dynamics in disordered media with experiment and theory}

\author{Andr\'es Montoya-Castillo}
\affiliation{Department of Chemistry, University of Colorado, Boulder, Boulder, Colorado, 80309, USA}

\author{Michael S. Chen}
\affiliation{Department of Chemistry, Stanford University, Stanford, California, 94305, USA}

\author{Sumana L. Raj}
\affiliation{Stanford PULSE Institute, SLAC National Accelerator Laboratory, Stanford University, Menlo Park, California 94025, USA}

\author{Kenneth A. Jung}
\affiliation{Department of Chemistry, Stanford University, Stanford, California, 94305, USA}

\author{Kasper S. Kjaer}
\affiliation{Stanford PULSE Institute, SLAC National Accelerator Laboratory, Stanford University, Menlo Park, California 94025, USA}

\author{Tobias Morawietz}
\affiliation{Department of Chemistry, Stanford University, Stanford, California, 94305, USA}

\author{Kelly J. Gaffney}
\email{kgaffney@slac.stanford.edu}
\affiliation{Stanford PULSE Institute, SLAC National Accelerator Laboratory, Stanford University, Menlo Park, California 94025, USA}

\author{Tim B. van Driel}
\email{timbvd@slac.stanford.edu}
\affiliation{Linac Coherent Light Source, SLAC National Accelerator Laboratory, Menlo Park, California 94025, USA}

\author{Thomas E. Markland}
\email{tmarkland@stanford.edu}
\affiliation{Department of Chemistry, Stanford University, Stanford, California, 94305, USA}

\date{\today}

\maketitle

\section{Derivation details}
\label{app:theory_deriv}

Here we derive a rigorous quantum mechanical framework in the linear response limit for predicting and interpreting experiments involving optically induced anisotropy combined with time-resolved scattering. We then specialize our theory to the INXS experiment by adopting a series of well benchmarked and thoroughly adopted approximations in the field of Raman spectroscopy and X-ray scattering to obtain expressions that are compatible with common treatments of polarizability (whether at the force field or electronic structure level) and molecular dynamics simulations as a means to sample the nuclear scattering pattern. Our approach is general and can be easily combined with different levels of theory. 

In a time-resolved X-ray scattering problem involving three incoming photons (two in the optical range and one in the X-ray range) and one scattered X-ray photon treated in the linear response limit, the measured signal will contain contributions from both the linear response arising from the incoming and scattered X-ray photons leading to a time-independent scattering pattern and the third-order response, 
\begin{equation}
    S^{\rm total}(\mathbf{q}, \tau) = S^{(1)}(\mathbf{q}) + S^{(3)}(\mathbf{q}, \tau).
\end{equation}
Here $S^{\rm total}(\mathbf{q}, \tau)$ is the total scattering signal arising from the experiment, $\tau$ is the waiting time between the optical photons (assumed to interact with the sample at approximately the same time, as is the case in INXS) and the X-ray interaction, $\mathbf{q}$ is the X-ray scattering wavevector, $S^{(1)}(\mathbf{q}) = S^{(1)}(q)$ is the isotropic time-independent linear X-ray scattering response, and $S^{(3)}(\mathbf{q}, \tau)$ is the the third-order time-dependent signal containing both isotropic and anisotropic contributions. The second-order response will be absent in systems with inversion symmetry, such as isotropic, disordered condensed phase systems away from an interface and all contributions higher than third order can be assumed to be negligible in the linear response limit for such an experiment \cite{Mukamel1995}. Since the third-order response entirely contains the novel dynamical information in the measurement, we isolate it by subtracting the first-order contribution from the total signal
\begin{equation}\label{eq:experimental-signal-third-order-relationship}
\begin{split}
    S^{(3)}(\mathbf{q}, \tau) &= S^{\rm total}(\mathbf{q}, \tau) - S^{(1)}(q)\\
    &\equiv \Delta S(\mathbf{q}, \tau).
\end{split}
\end{equation}
This process of subtracting the time-independent (arising from the impulsive limit of X-ray scattering, see below), isotropic linear X-ray scattering contribution is consistent with manipulations performed on previous experiments \cite{Lorenz2010,Biasin2016,Ki2021,Kim2020} and we have chosen our notation for the remainder, $\Delta S(\mathbf{q}, \tau)$ to mirror the notation used in these studies. Hence, the rest of this section centers on obtaining a rigorous quantum mechanical expression for the third-order response, $\Delta S(\mathbf{q}, \tau)$, that can serve as the starting point for a series of approximations to render the theoretical expression for our time-resolved X-ray scattering signal compatible with existing molecular simulation tools. 

To obtain an expression for $\Delta S(\mathbf{q}, \tau)$, we start from a modified expression for the time-resolved X-ray diffraction signal in Eq.~E7 in Appendix E of Ref.~\onlinecite{Tanaka2001},
\begin{equation}\label{eq:3rd-order-response-density-matrix}
\begin{split}
    \Delta& S(\mathbf{q}, \tau) = \int dt\ \langle \langle \mathcal{N}_s|\rho^{(3)}(t)\rangle \rangle\\
    &\propto \frac{2(\mathbf{e}_{\rm xr} \cdot \mathbf{e}_{\rm s})^2|E_{\rm s}|^2}{\omega_{\rm opt}^2 \omega_{\rm xr}^2 \omega_{\rm s}^2} \mathrm{Re} \int d\mathbf{r}
    \int d\mathbf{r}_1 \int d\mathbf{r}_2 \int d\mathbf{r}_3 \int dt \int dt_1 \int dt_2 \int dt_3\\
    & \times E^*_{\rm xr}(t-t^3)E_{\rm xr}(t) e^{-i\mathbf{q} \cdot (\mathbf{r} - \mathbf{r}_3)} e^{i \Delta \omega t_3}\\
    & \times \Big\{ E_{\rm opt}(t - t_3 - t_2 - t_1 + \tau) E_{\rm opt}^*(t-t_3-t_2+\tau)  \Big[R_{1}(\mathbf{R},\mathbf{t}) + R_{4}(\mathbf{R},\mathbf{t}) \Big]  e^{i\omega_{\rm opt} t_1 } \Big\}\\
    & + \Big\{ E^*_{\rm opt}(t - t_3 - t_2 - t_1 + \tau) E_{\rm opt}(t-t_3-t_2+\tau)  \Big[R_{2}(\mathbf{R},\mathbf{t}) + R_{3}(\mathbf{R},\mathbf{t}) \Big]  e^{-i\omega_{\rm opt} t_1 } \Big\},
\end{split}
\end{equation}
where $\mathbf{R} \equiv \{ \mathbf{r}, \mathbf{r}_3, \mathbf{r}_2,\mathbf{r}_1 \}$ and $\mathbf{t} \equiv \{ t_3, t_2, t_1 \}$, $E_{\rm opt}(t)$ and $E_{\rm xr}(t)$ are the electric field envelopes for the optical and incoming X-ray pulses, respectively, $\mathbf{q} \equiv \mathbf{k}_{\rm s} - \mathbf{k}_{\rm xr}$ is the X-ray scattering wavevector and $\Delta \omega \equiv \omega_{\rm s} - \omega_{\rm xr}$ is the difference in frequency between the incoming and scattered X-ray pulses, indicating the energy loss. This expression adopts the \textit{dipole approximation}, which sets the wavevector for the optical field as approximately zero (i.e., $\mathbf{k}_{\rm opt} = 2\pi \mathbf{e}_{\rm opt} / \lambda_{\mathrm{opt}} \approx 0$, where $\mathbf{e}_{\rm opt}$ is the unit vector which determines the direction of propagation for the optical field). This expression also assumes the validity of the \textit{slowly varying field envelope approximation}, which is valid when the field amplitudes change more slowly than the optical periods and thus replaces the time derivative of the vector potential with the electric field, $\dot{\mathbf{A}} \rightarrow - \mathbf{E}$ in the original formulation based on the minimal coupling Hamiltonian \cite{Craig1984}. In addition, the expression in Eq.~\ref{eq:3rd-order-response-density-matrix} assumes the \textit{rotating wave approximation} with respect to the optical and X-ray pulses.The modification to Eq.~E7 in Appendix E of Ref.~\onlinecite{Tanaka2001} in Eq.~\ref{eq:3rd-order-response-density-matrix} above centers on the prefactor, which accounts for the dependence of the signal on: the frequency of the scattered X-ray, $\omega_{\rm s}$, the inner product of the polarization of the incoming and scattered X-ray pulses, $(\mathbf{e}_{\rm xr} \cdot \mathbf{e}_{\rm s})^2$, and of the power of the scattered electric field $|E_{\rm s}|^2$, which arise from considering the vector nature of the X-ray electric fields and the application of the slowly varying approximation. Finally, $R_i$ is the i$th$ response function corresponding to the four distinct Liouville pathways, for $i \in \{1, 2, 3, 4 \}$. These response functions take the following forms, 
\begin{subequations}\label{eq:response-equations}
\begin{align}
    R_1 &= \mathrm{Tr}\Big[ \hat{j}_{\lambda}(\mathbf{r}_2, t_1)\hat{\sigma}(\mathbf{r}_3, t_1 + t_2) \hat{\sigma}(\mathbf{r}, t_1+t_2 + t_3) \hat{j}_{\lambda}(\mathbf{r}_1, 0) \rho\Big],\\
    R_2 &= \mathrm{Tr}\Big[\hat{j}_{\lambda}(\mathbf{r}_1, 0)\hat{\sigma}(\mathbf{r}_3, t_1 + t_2) \hat{\sigma}(\mathbf{r}, t_1+t_2 + t_3) \hat{j}_{\lambda}(\mathbf{r}_2, t_1) \rho\Big],\\
    R_3 &= - \mathrm{Tr}\Big[\hat{j}_{\lambda}(\mathbf{r}_1, 0)\hat{j}_{\lambda}(\mathbf{r}_2, t_1) \hat{\sigma}(\mathbf{r}_3, t_1 + t_2) \hat{\sigma}(\mathbf{r}, t_1+t_2 + t_3) \rho\Big],\\
    R_4 &= -\mathrm{Tr}\Big[\hat{\sigma}(\mathbf{r}_3, t_1 + t_2) \hat{\sigma}(\mathbf{r}, t_1+t_2 + t_3) \hat{j}_{\lambda}(\mathbf{r}_2, t_1) \hat{j}_{\lambda}(\mathbf{r}_1, 0) \rho\Big],
\end{align}
\end{subequations}
where $\rho$ is the equilibrium (canonical) density operator for the material being probed in the absence of the light field, and 
\begin{subequations}
\begin{align}
    \hat{\mathbf{j}}(\mathbf{r}) &= \frac{e\hbar}{2mi} \Big[\hat{\psi}^{\dagger}(\mathbf{r}) \boldsymbol{\nabla} \hat{\psi}(\mathbf{r}) - [\boldsymbol{\nabla} \hat{\psi}^{\dagger}(\mathbf{r})] \hat{\psi}(\mathbf{r}) \Big],\label{eq:current-operator}\\
    \hat{\sigma(\mathbf{r})} &= \hat{\psi}^{\dagger}(\mathbf{r})\hat{\psi}(\mathbf{r}), \label{eq:charge-density-operator}
\end{align}
\end{subequations}
are the current and charge density operators, respectively. The $\lambda$ component of the current operator, $\hat{j}_{\lambda}(\mathbf{r})$, in the response functions in Eq.~\ref{eq:response-equations}, arises from the inner product of $\hat{\mathbf{j}}(\mathbf{r})$ and the polarization vector, $\boldsymbol{\epsilon}_{\rm opt}$, of the electric field of the optical pulse, $\mathbf{E}_{\rm opt}(t)$, responsible for this interaction. We take the electric field vector to be given by 
\begin{equation} \label{eq:optical-pulse-electric-field}
    \mathbf{E}_{\rm opt}(t) = \boldsymbol{\epsilon}_{\rm opt}E_{\rm opt}(t) e^{i(\mathbf{k}\cdot \mathbf{r} - \omega t) }+ \boldsymbol{\epsilon}_{\rm opt}^*E^*_{\rm opt}(t) e^{-i(\mathbf{k}\cdot \mathbf{r} - \omega t) }.
\end{equation}
Using this notation, the $\lambda$ component of the current vector operator is
\begin{equation}
    \hat{j}_{\lambda}(\mathbf{r}) \equiv \boldsymbol{\epsilon}_{\rm opt} \cdot \hat{\mathbf{j}}(\mathbf{r}).
\end{equation}

Since we are particularly interested in optically induced anisotropy in scattering experiments, it is worthwhile to move to the (spatial) Fourier representation, which yields the following expression for the time-resolved scattering signal
\begin{equation}\label{eq:trxd-signal-qm-fourier-space}
\begin{split}
    \Delta S(\mathbf{q}, \tau) &= \frac{2(\mathbf{e}_{\rm xr} \cdot \mathbf{e}_{\rm s})^2|E_{\rm s}|^2}{\omega_{\rm opt}^2 \omega_{\rm xr}^2 \omega_{\rm s}^2} \mathrm{Re}  \int dt \int dt_1 \int dt_2 \int dt_3\  E_{\rm xr}(t-t_3)E_{\rm xr}(t) e^{i \Delta \omega t_3}\\
    &\qquad \times \Big\{ e^{i\omega_{\rm opt} t_1}E_{\rm opt}(t - t_3-t_2-t_1)E_{\rm opt}^*(t - t_3-t_2)  \Big[ R_1(\mathbf{q}, \mathbf{t}) + R_4(\mathbf{q}, \mathbf{t}) \Big]\\
    &\qquad \qquad  e^{-i\omega_{\rm opt} t_1}E^*_{\rm opt}(t - t_3-t_2-t_1)E_{\rm opt}(t - t_3-t_2) \Big[ R_2(\mathbf{q}, \mathbf{t}) + R_3(\mathbf{q}, \mathbf{t}) \Big] \Big\},
\end{split}
\end{equation}
where the response functions take the forms,
\begin{subequations}\label{eq:response-functions-q}
\begin{align}
    \tilde{R}_1 &= \mathrm{Tr}\Big[ \hat{j}_{\lambda}(t_1)\hat{\sigma}(-\mathbf{q}, t_1 + t_2) \hat{\sigma}(\mathbf{q}, t_1+t_2 + t_3) \hat{j}_{\lambda}(0) \rho\Big],\\
    \tilde{R}_2 &= \mathrm{Tr}\Big[\hat{j}_{\lambda}(0)\hat{\sigma}(-\mathbf{q}, t_1 + t_2) \hat{\sigma}(\mathbf{q}, t_1+t_2 + t_3) \hat{j}_{\lambda}(t_1) \rho\Big],\\
    \tilde{R}_3 &= - \mathrm{Tr}\Big[\hat{j}_{\lambda}(0)\hat{j}_{\lambda}(t_1) \hat{\sigma}(-\mathbf{q}, t_1 + t_2) \hat{\sigma}(\mathbf{q}, t_1+t_2 + t_3) \rho\Big],\\
    \tilde{R}_4 &= -\mathrm{Tr}\Big[\hat{\sigma}(-\mathbf{q}, t_1 + t_2) \hat{\sigma}(\mathbf{q}, t_1+t_2 + t_3) \hat{j}_{\lambda}(t_1) \hat{j}_{\lambda}(0) \rho\Big].
\end{align}
\end{subequations}
Here we have adopted the following expression for the Fourier space representation of a function of coordinates, 
\begin{equation}
    \mathcal{O}(\mathbf{q}) = \int d\mathbf{r}\ e^{-i\mathbf{q} \cdot \mathbf{r}} \mathcal{O}(\mathbf{r}).
\end{equation}
To arrive at Eqs.~\ref{eq:response-functions-q}, we have also performed the integrations over $\mathbf{r}_1$ and $\mathbf{r}_2$. Since the dipole approximation with respect to the optical field sets $\mathbf{k}_{\rm opt} = 0$, the Fourier representation of these spatial variables is taken at $\mathbf{q}_i = 0$ and thus do not appear in Eqs.~\ref{eq:response-functions-q}. Hence, the modified current operator takes the form,
\begin{equation}
    \hat{j}_{\lambda} \equiv \int d\mathbf{r}\ \hat{j}_{\lambda}(\mathbf{r}).
\end{equation}

Up to this point, our discussion has been generally applicable to third-order scattering experiments that combine optical excitation with X-ray scattering and has only employed established approximations (e.g., the dipole approximation for the optical pulse and the slowly varying field envelope approximation). To specialize our discussion to the INXS experiment, we now turn to the details of the INXS setup and exploit several details from the experiment. In particular:
\begin{enumerate}
    \item The X-ray scattering in the INXS experiment is elastic, meaning that $\Delta \omega = 0$. While there may be incoherent contributions that go beyond the elastic limit, these can be expected to be small and only become a concern on the attosecond scale \cite{Dixit2012}.
    
    \item The duration of the X-ray interaction with matter is extremely short, permitting it to be approximated as instantaneous and resulting in the impulsive limit for the X-rays, which arrive at a time delay $\tau$ after the interaction of matter with the optical pulses, $E_{\rm xr}(t) = E_{\rm xr} \delta(t - \tau)$. This allows us to explicitly perform the integrations over $t$ and $t_3$ in Eq.~\ref{eq:trxd-signal-qm-fourier-space}, leading to the following replacements $t \rightarrow \tau$ and $t_3 \rightarrow 0$ in Eqs.~\ref{eq:trxd-signal-qm-fourier-space} and \ref{eq:response-functions-q}. 
    
    \item The system is in thermal equilibrium with a thermal energy of $\beta = [k_BT]^{-1}$ assumed to be insufficient to populate electronically excited states. This means the density operator for the system before interaction with the light is given by $\rho = \ket{g} \rho_{\rm nuc}^{(g)} \bra{g}$, where $\rho_{\rm nuc}^{(g)}$ is the canonical density operator for the nuclei on the ground state potential energy surface and $\ket{g}$ is the ground electronic state. In turn, this means that only the ground state term in the trace over the electronic degrees of freedom, $\mathrm{Tr}_{\rm elec}[...] = \braket{g|...|g} + \sum_{\{e\}} \braket{e|...|e}$, survives for each of the response functions, i.e., $R_{i} = \mathrm{Tr}_{\rm nuc}[\braket{g| ABCD | g}\rho_{\rm nuc}^{(g)}]$ where $A, B, C, D \in  \{\hat{j}, \hat{\sigma} \}$. 
    
    \item The Raman and X-ray pulses are off-resonance. This implies that the light-matter interactions do not generate electronic excitations and, hence response functions that contain isolated current operators, $\hat{j}$, which cause electronic excitations, provide no contributions. In other words, the only response functions that survive are those where the current operators, $\hat{j}$, are adjacent to each other, i.e.,  $R_3$ and $R_4$. 
    
    \item Since the optical pulse leads to a Raman interaction, the two photons interact with matter in quick succession, which are separated by a waiting time of $t_1$. Amplitude variations in the electric field envelope for the optical pulse, $E_{\rm opt}(t)$, during $t_1$ can be expected to be small, and hence negligible \cite{Cho1993,Yan1991,Tanimura1993}. This means that $E_{\rm opt}(s \pm t_1) \approx E_{\rm opt}(s)$, where $s$ is some particular combination of time variables. 
    
\end{enumerate}

Employing these approximations allows one to rewrite Eq.~\ref{eq:trxd-signal-qm-fourier-space} 
\begin{equation}\label{eq:trxd-signal-qm-expt-approx}
\begin{split}
    \Delta S(\mathbf{q}, \tau) &= \mathcal{N}^{\rm qm} \mathrm{Re}  \int dt_1 \int dt_2\  \Big\{ e^{i\omega_{\rm opt} t_1}E_{\rm opt}(\tau -t_2 -t_1)E_{\rm opt}^*(\tau - t_2) R_4(\mathbf{q}, \mathbf{t}) \\
    &\qquad \qquad \qquad \qquad  + e^{-i\omega_{\rm opt} t_1}E^*_{\rm opt}(\tau -t_2 -t_1)E_{\rm opt}(\tau -t_2)  R_3(\mathbf{q}, \mathbf{t})  \Big\}\\
    &\approx \mathcal{N}^{\rm qm} \mathrm{Re}  \int dt_1 \int dt_2\ |E_{\rm opt}(\tau - t_2)|^2  \Big\{ e^{i\omega_{\rm opt} t_1} R_4(\mathbf{q}, \mathbf{t}) +e^{-i\omega_{\rm opt} t_1} R_3(\mathbf{q}, \mathbf{t})  \Big\}\\
\end{split}
\end{equation}
where
\begin{equation}\label{eq:prelim-proportionality-factor}
    \mathcal{N}^{\rm qm} = \frac{(\mathbf{e}_{\rm xr} \cdot \mathbf{e}_{\rm s})^2|E_{\rm s}|^2 |E_{\rm xr}|^2}{\omega_{\rm opt}^2 \omega_{\rm xr}^2 \omega_{\rm s}^2},
\end{equation}
and the response functions take the forms,
\begin{subequations}\label{eq:response-functions-q-spatial-FT}
\begin{align}
    \tilde{R}_3 &= - \mathrm{Tr}_{\rm nuc}\Big[\braket{g|\hat{j}_{\lambda}(0)\hat{j}_{\lambda}(t_1)|g} \hat{\Sigma}(\mathbf{q}, t_1+t_2) \rho_{\rm nuc}^{(g)}\Big],\\
    \tilde{R}_4 &= -\mathrm{Tr}_{\rm nuc}\Big[\hat{\Sigma}(\mathbf{q}, t_1+t_2) \braket{g|\hat{j}_{\lambda}(t_1) \hat{j}_{\lambda}(0)|g} \rho_{\rm nuc}^{(g)}\Big].
\end{align}
\end{subequations}
Here we have defined the charge density scattering operator on the ground potential energy surface, $\hat{\Sigma}(\mathbf{q})$, as
\begin{equation}\label{eq:charge-density-scattering-operator}
    \hat{\Sigma}(\mathbf{q}) \equiv \hat{\sigma}_{g}(-\mathbf{q}) \hat{\sigma}_{g}(\mathbf{q}).
\end{equation}

\subsection{Treatment of X-ray scattering}
\label{ssec:xray-scattering}

Central to the X-ray scattering component of the INXS experiment and our formulation of its signal is the charge density scattering operator on the ground potential energy surface, $\hat{\Sigma}(\mathbf{q})$, in Eq.~\ref{eq:charge-density-scattering-operator}. Here we outline how we use a series of approximations commonly used in X-ray scattering to obtain an expression that is compatible with direct simulation. We begin by rewriting the expression for $\hat{\Sigma}(\mathbf{q})$ in terms of electron creation and annihilation operators, $\hat{c}_{\alpha, l}^{\dagger}$ and $\hat{c}_{\alpha, l}$, where the index $l$ labels atoms and $\alpha$ the electron in the $l$th atom in the system, 
\begin{equation}\label{eq:charge-density-scattering-operator-IAM}
\begin{split}
    \hat{\sigma}(\mathbf{q}) &= \int_{\mathrm{V}} d\mathbf{r}\  e^{-i\mathbf{q} \cdot \mathbf{r}}\hat{\psi}^{\dagger}(\mathbf{r}) \hat{\psi}(\mathbf{r})\\
    &= \sum_{\alpha, \alpha'} \sum_{l, l'} \hat{c}_{\alpha', l'}^{\dagger}\hat{c}_{\alpha, l} \int_{\mathrm{V}} d\mathbf{r}\ e^{-i\mathbf{q} \cdot \mathbf{r}}\phi_{\alpha}(\mathbf{r} - \mathbf{R}_l) \phi_{\alpha '}^*(\mathbf{r} - \mathbf{R}_{l'})\\
    &\approx \sum_{l} \sum_{\alpha, \alpha'}  \hat{c}_{\alpha', l}^{\dagger}\hat{c}_{\alpha, l} \int_{\mathrm{V}} d\mathbf{r}\ e^{-i\mathbf{q} \cdot \mathbf{r}}\phi_{\alpha}(\mathbf{r} - \mathbf{R}_l) \phi_{\alpha '}^*(\mathbf{r} - \mathbf{R}_{l})\\
    &= \sum_{l} \sum_{\alpha, \alpha'} \hat{c}_{\alpha', l}^{\dagger}\hat{c}_{\alpha, l} e^{-i\mathbf{q} \cdot \mathbf{R}_l} \eta_{l, \alpha, \alpha'}(\mathbf{q}),
\end{split}
\end{equation}
where we have used the expression for the charge density operators, Eq.~\ref{eq:charge-density-operator}, in terms of electron field operators, 
\begin{subequations}
\begin{align}
    \hat{\psi}(\mathbf{r}) &= \sum_{\alpha, l} \phi_{\alpha}(\mathbf{r} - \mathbf{R}_l)\hat{c}_{\alpha, l},\\
    \hat{\psi}(\mathbf{r})^{\dagger} &= \sum_{\alpha, l} \phi_{\alpha}^*(\mathbf{r} - \mathbf{R}_l)\hat{c}_{\alpha, l}^{\dagger},
\end{align}
\end{subequations}
and
\begin{equation}
    \eta_{l, \alpha, \alpha'}(\mathbf{q}) \equiv \int_{\mathrm{V}} d\mathbf{r}\ e^{-i\mathbf{q} \cdot \mathbf{r}}\phi_{\alpha}(\mathbf{r} ) \phi_{\alpha '}^*(\mathbf{r}).
\end{equation}
Here, $\{\phi_{\alpha}^*(\mathbf{r} - \mathbf{R}_l) \}$ is an atom-centered localized basis, where $\mathbf{R}_l$ denotes the position of the $l$th atom. To arrive at the penultimate line of Eq.~\ref{eq:charge-density-scattering-operator-IAM}, we have adopted the \textit{independent atom model} (IAM) limit \cite{Moller2012,Helliwell1997} which assumes that the electron density --- which is responsible for the scattering of X-rays --- can be written as a sum of free atomic densities, i.e., the electron clouds of all atoms are well localized and scatter X-rays independently. This approximation has been thoroughly tested in the context of structure determination from X-ray scattering experiments and is generally considered to be sufficiently accurate \cite{Coppens1992,Ben-Nun1997,Cao1998,Rozgonyi2005,Dohn2015}. Nevertheless, the approach that we outline here is general and compatible with more sophisticated treatments of the scattering operator \cite{MorenoCarrascosa2017,Northey2014,Northey2016,MorenoCarrascosa2019}.

Mathematically, the IAM considers the overlap between two atom-centered basis functions as being negligible unless they lie on the same atom,
\begin{equation}
    \phi_{\alpha}(\mathbf{r} - \mathbf{R}_l) \phi_{\alpha '}^*(\mathbf{r} - \mathbf{R}_{l'}) = \delta_{l, l'} \phi_{\alpha}(\mathbf{r} - \mathbf{R}_l) \phi_{\alpha '}^*(\mathbf{r} - \mathbf{R}_{l}),
\end{equation}
thus allowing one to significantly simplify the expressions for the charge density operator based on the assumption that electrons do not jump from chemical species to chemical species (in this case, atoms or molecules).

Within the IAM, the contribution arising from cross terms $\alpha \neq \alpha '$ in Eq.~\ref{eq:charge-density-scattering-operator-IAM} will be zero. This approximation is consistent with the condition that the incoming X-rays are off-resonance and therefore do not cause electronic excitations (see item \# 4 in the list of experimental details that inform our construction of the theory). In this limit, the Fourier space charge density operator matrix element corresponding to the electronic ground state takes the form, 
\begin{equation}\label{eq:charge-density-scattering-operator-IAM-no-electronic-excitations}
\begin{split}
    \hat{\sigma}_{g}(\mathbf{q}) &\equiv \bra{g}\hat{\sigma}(\mathbf{q})\ket{g}\\
    &= \sum_{l} e^{i\mathbf{q} \cdot \mathbf{R}_l} \sum_{\alpha \in \mathrm{occ}}\eta_{l, \alpha, \alpha}(\mathbf{q})\\
    &= \sum_{l} e^{i\mathbf{q} \cdot \mathbf{R}_l} F_l(\mathbf{q}; \{ \mathbf{R} \}),
\end{split}
\end{equation}
where  
\begin{equation}\label{eq:atomic-form-factor}
\begin{split}
    F_l(\mathbf{q}; \{ \mathbf{R} \}) &= \sum_{\alpha} \eta_{l, \alpha, \alpha}^{occ}(\mathbf{q}; \{ \mathbf{R} \})\\
    &= \sum_{\alpha \in \mathrm{occ}} \int_{\mathbf{r} \in l} d\mathbf{r}\ e^{-i\mathbf{q} \cdot \mathbf{r}}|\phi_{\alpha}(\mathbf{r; \{ \mathbf{R} \}} )|^2 
\end{split}
\end{equation}
is known as the atomic form factor \cite{Als-Nielsen2011}. Generally, the molecular form factor depends not only on the electronic coordinates but also parametrically on the nuclear coordinates $\mathbf{R}$. 

The IAM also makes two additional approximations regarding the symmetries that the atomic form factors obey. These assert that atomic form factors (i) do not depend strongly on the global nuclear configuration of the system, i.e., $F_{l}(\mathbf{q}; \{ \mathbf{R} \}) \approx F_{l}(\mathbf{q})$, and (ii) are approximately spherically symmetric, $F_{l}(\mathbf{q}) = F_{l}(q)$, which implies that  the form factor is real, $F_{l}(q) = F^{*}_{l}(-q) = F_{l}(-q)$ \cite{Palinkas1973,Soper2007}. The approximations at the heart of the IAM form the basis of much work in X-ray scattering \cite{Moller2012,Helliwell1997} and can be easily relaxed in future implementations of the INXS theory, should it prove necessary. With these approximations, noting that $F(\mathbf{q})^* = F(-\mathbf{q})$, and substituting Eqs.~\ref{eq:charge-density-scattering-operator-IAM-no-electronic-excitations} and \ref{eq:atomic-form-factor} into Eq.~\ref{eq:charge-density-scattering-operator}, one obtains the following expression for charge density scattering operator on the ground potential energy surface,
\begin{equation}\label{eq:scattering-operator-IAM}
\begin{split}
    \hat{\Sigma}(\mathbf{q}) &\approx \sum_{l, l'} e^{-i\mathbf{q} \cdot (\mathbf{R}_l - \mathbf{R}_{l'})} F_l(q)F_{l'}(q)\\
    &= \sum_{l} F_l^2(q) + \sum_{l \neq l'} e^{-i\mathbf{q} \cdot (\mathbf{R}_l - \mathbf{R}_{l'})} F_l(q)F_{l'}(q).
\end{split}
\end{equation}
Note that in the second line of Eq.~\ref{eq:scattering-operator-IAM}, we have separated the contributions due to self-scattering ($l = \l'$) and from scattering from pairs of distinct atoms ($l \neq \l'$). 

\subsection{Treatment of the Raman interaction}
\label{ssec:Raman-interaction}

We now invoke a commonly made approximation in multidimensional spectroscopies involving Raman pulses which asserts that the two optical pulses that compose a Raman interaction are short compared with the timescales associated with molecular motions (albeit not necessarily electronic motions), which allows one to ignore nuclear dynamics during the short timescale separating the two optical pulses. Since the INXS experiment employs a Raman interaction to optically induce anisotropy in the X-ray scattering and this is a commonly adopted and thoroughly benchmarked approximation in Raman spectroscopy \cite{Cho1993,Yan1991,Tanimura1993}, one can expect this to be a reasonable approximation for many systems, especially those where librations can be much faster (e.g., water). We apply this approximation on the $R_3$ and $R_4$ response functions in Eq.~\ref{eq:response-functions-q} with respect to the waiting time between the two pulses, $t_1$. In particular, since we are primarily concerned with X-ray scattering from the nuclei in the INXS experiment, one can neglect $t_1$ in the argument of the charge density scattering operator, $\hat{\Sigma}(\mathbf{q})$, whose time variation arises primarily from nuclear rather than electronic dynamics, leading to
\begin{subequations}\label{eq:response-functions-q-neglected-nuclear-dynamics-t1}
\begin{align}
    \tilde{R}_3 &= - \mathrm{Tr}_{\rm nuc}\Big[\braket{g|\hat{j}_{\lambda}(0)\hat{j}_{\lambda}(t_1)|g} \hat{\Sigma}(\mathbf{q}, t_2) \rho_{\rm nuc}^{(g)}\Big],\\
    \tilde{R}_4 &= -\mathrm{Tr}_{\rm nuc}\Big[\hat{\Sigma}(\mathbf{q}, t_2) \braket{g|\hat{j}_{\lambda}(t_1) \hat{j}_{\lambda}(0)|g} \rho_{\rm nuc}^{(g)}\Big].
\end{align}
\end{subequations}
Application of this approximation to $\braket{g|\hat{j}_{\lambda}(0)\hat{j}_{\lambda}(t_1)|g}$ and $\braket{g|\hat{j}_{\lambda}(t_1)\hat{j}_{\lambda}(0)|g}$ requires care. Specifically, while the neglect of nuclear dynamics in the case of $\hat{\Sigma}(\mathbf{q})$ resulted in setting $t_1 \rightarrow 0$, since its contribution to the time-resolved X-ray scattering signal is dominated by the dynamics of the nuclei rather than those of the electrons, in the case of the charge current, we need to consider the electronic dynamics subject to frozen nuclei (on the timescales of $t_1$) \cite{Yan1991,Cho1993,Tanimura1993}, leading to the following expressions for the $R_3$ and $R_4$ response functions, 
\begin{subequations}\label{eq:response-functions-R3-R4}
\begin{align}
    \tilde{R}_3 &\equiv \int dt_1 \   e^{-i\omega_{\rm opt} t_1} R_3(\mathbf{q}, \mathbf{t}) \nonumber \\
    &= - \sum_{e} \mathcal{G}_{eg}(\omega_{\rm opt}) \mathrm{Tr}_{\rm nuc}\Big[\hat{j}_{ge}(0)\hat{j}_{eg}(0) \hat{\Sigma}_g(\mathbf{q}, t_2) \rho_g\Big], \label{eq:response-functions-R3}\\
    \tilde{R}_4 &\equiv \int dt_1 \   e^{i\omega_{\rm opt} t_1} R_4(\mathbf{q}, \mathbf{t}) \nonumber \\
    &= -\sum_{e} \mathcal{G}_{ge}(-\omega_{\rm opt}) \mathrm{Tr}_{\rm nuc}\Big[\hat{\Sigma}_g(\mathbf{q}, t_2) \hat{j}_{ge}(0) \hat{j}_{eg}(0) \rho_g\Big],\label{eq:response-functions-R4}
\end{align}
\end{subequations}
where $\mathcal{G}_{ge}(\omega)$ is the electronic Green's function corresponding to the electronic Hamiltonian evolution $\hat{j}_{\lambda}(t_1) = e^{iH_{\rm elec}t/\hbar}\hat{j}e^{-iH_{\rm elec}t/\hbar}$, where $H_{\rm elec}(\mathbf{q}) = \ket{g} E_g(\mathbf{R}) \bra{g} + \sum_{e} \ket{e}E_e(\mathbf{R})\bra{e}$, $E_g(\mathbf{R})$ is the potential energy surface for the electronic ground state and $\{E_{e}(\mathbf{R}) \}$ is the set of potential energy surfaces for the electronic excited states, which depend parametrically on the nuclear coordinates $ \{ \mathbf{R} \}$ of the sample. Note that this evolution neglects the time evolution of the nuclei over $t_1$, consistent with the approximation above. The electronic Green's function takes the form,
\begin{equation}\label{eq:greens-function-eg-fourier}
\begin{split}
    \mathcal{G}_{eg}(\omega) &= \int_0^{\infty} dt\ e^{i\omega t} e^{-i\omega_{eg} t}\\
    &= \int_{-\infty}^{\infty} dt\ \theta(t) e^{i(\omega - \omega_{eg}) t}\\
    &= - \lim_{\eta \rightarrow 0^{+}}\int_{-\infty}^{\infty} dt\ \int \frac{d\Omega}{2\pi i} \frac{e^{-i[\Omega - (\omega - \omega_{eg})] t}}{\Omega + i\eta} = - \lim_{\eta \rightarrow 0^{+}} \frac{2\pi}{2\pi i} \frac{1}{\omega - \omega_{eg} + i\eta} = \frac{i}{\omega - \omega_{eg}},
\end{split}
\end{equation}
where we have used the following expression for the Heaviside function,
\begin{equation} \label{eq:heaviside-function-fourier}
\begin{split}
    \theta(t) &= - \lim_{\eta \rightarrow 0^{+}} \int \frac{d\Omega}{2\pi i} \frac{e^{-i\Omega t}}{\Omega + i\eta},\\
    &= \lim_{\eta \rightarrow 0^{+}} \int \frac{d\Omega}{2\pi i} \frac{e^{i\Omega t}}{\Omega - i\eta}.
\end{split}
\end{equation}
Using Eqs.~(\ref{eq:greens-function-eg-fourier}) and (\ref{eq:heaviside-function-fourier}), one can show that 
\begin{equation}\label{eq:greens-function-ge-fourier}
\begin{split}
    \mathcal{G}_{ge}(-\omega) &= \int_0^{\infty} dt\ e^{-i\omega t} e^{+i\omega_{eg} t}\\
    &= \int_{-\infty}^{\infty} dt\ \theta(t) e^{-i(\omega - \omega_{eg}) t}\\
    &= \lim_{\eta \rightarrow 0^{+}}\int_{-\infty}^{\infty} dt\ \int \frac{d\Omega}{2\pi i} \frac{e^{-i[\Omega - (\omega - \omega_{eg})] t}}{\Omega - i\eta} = \lim_{\eta \rightarrow 0^{+}} \frac{2\pi}{2\pi i} \frac{1}{\omega - \omega_{eg} - i\eta} = - \frac{i}{\omega - \omega_{eg}},
\end{split}
\end{equation}
which implies that 
\begin{equation}\label{eq:relation-greens-function-fourier-eg-ge}
    \mathcal{G}_{eg}(\omega) = - \mathcal{G}_{ge}(-\omega).
\end{equation}

Since the materials being studied in the experiment is not a metal, we take the static limit $\omega_{\rm opt} \rightarrow 0$ \cite{Luber2014,Long2002} and recover the expression for the polarizability tensor,
\begin{equation}\label{eq:poralizability-tensor-static-limit}
\begin{split}
    \hat{\alpha}_{ab} &= 2 \sum_{e} \frac{\hat{j}^{(a)}_{ge}\hat{j}^{(b)}_{eg}}{\omega_{eg}}\\
    &= -2i \lim_{\omega_{\rm opt} \rightarrow 0}  \sum_{e} \hat{j}^{(a)}_{ge}\hat{j}^{(b)}_{eg} \mathcal{G}_{eg}(\omega),
\end{split}
\end{equation}
where $a$ and $b$ are the polarizations of the electric fields which gives rise to the transition dipole operators. This implies that 
\begin{equation}
    \lim_{\omega_{\rm opt} \rightarrow 0}  \sum_{e} \hat{j}^{(a)}_{ge}\hat{j}^{(b)}_{eg} \mathcal{G}_{eg}(\omega) = \frac{i}{2} \alpha_{ab}.
\end{equation}

In our INXS experiment, the Raman excitation arises from a single light source, which implies that the two optical pulses generating the Raman interaction have the same polarization. Setting the lab frame such that the direction of propagation of the optical pulse is in the positive $z$ direction (i.e., $\mathbf{k}_{\rm opt} = \hat{\mathbf{z}} k$), then the polarization of the optical pulses lies on the $x$-$y$. The most general expression for such a polarization is given by
\begin{equation}\label{eq:general-polarization-on-xy-plane}
    \boldsymbol{\epsilon}_{\rm opt} = \cos(\phi)\hat{\mathbf{x}} + e^{i\delta}\sin(\phi)\hat{\mathbf{y}},
\end{equation}
where $\hat{\mathbf{x}}$ and $\hat{\mathbf{y}}$ are unit vectors in the $x$ and $y$ directions, respectively. When the phase, $\delta = 0$, the angle $\phi$ determines the angle from the $x$-axis of the linear polarization of the optical pulse. The phase $\delta$ can be used to distinguish linear ($\delta = 0, \pi$), circular, $\delta = \pm \frac{\pi}{2}$, and elliptical (all other values of $\delta$). This implies that the $\lambda$-component of the current operator selects a linear combination of the $x$ and $y$ components
\begin{equation}\label{eq:general-current-components-on-xy-plane}
   \hat{j}_{\lambda} = \cos(\phi)\hat{j}^{(x)} + e^{i\delta}\sin(\phi)\hat{j}^{(y)},
\end{equation}
and the $\lambda,\lambda$ component of the polarizability tensor is a linear combination of the diagonal and off-diagonal components,
\begin{equation} \label{eq:inxs-elements-of-polarizability-tensor}
    \hat{\tilde{\alpha}}(\mathbf{R}) = \cos^2(\phi)\hat{\alpha}_{xx}(\mathbf{R}) + \sin^2(\phi)\hat{\alpha}_{yy}(\mathbf{R}) + \cos(\delta)\sin(2\phi)[\hat{\alpha}_{xy}(\mathbf{R}) + \hat{\alpha}_{yx}(\mathbf{R})],
\end{equation}
This means that the current operator contains contributions in the $x$- and $y$-directions:
\begin{equation}
    \hat{j} \rightarrow \cos(\phi)\hat{j}^{(x)} + \sin(\phi)\hat{j}^{(y)},
\end{equation}
implying that our INXS experiment requires a linear combination of the different elements of the polarizability tensor,
\begin{equation} \label{eq:inxs-elements-of-polarizability-tensor}
    \hat{\tilde{\alpha}}(\mathbf{R}) = \cos^2(\phi)\hat{\alpha}_{xx}(\mathbf{R}) + \sin^2(\phi)\hat{\alpha}_{yy}(\mathbf{R}) + \cos(\phi)\sin(\phi)[\hat{\alpha}_{xy}(\mathbf{R}) + \hat{\alpha}_{yx}(\mathbf{R})],
\end{equation}
where we have added the implicit parametric dependence of the polarizability tensor on the coordinates of all the nuclei in the system for emphasis. In our experiment, we use linearly polarized light, which implies that we take $\delta = 0$ in Eq.~\ref{eq:inxs-elements-of-polarizability-tensor}. However, our treatment also delineates how using other polarizations, one may obtain distinct INXS signals. For instance, when using circularly polarized light, $\delta = \pm \frac{\pi}{2}$, the term containing the off-diagonal polarizability tensors would disappear. In such a case, the INXS experiment would only yield an anisotropic signal when the different diagonal components of the polarizability tensor are different, i.e., $\alpha_{j,j} \neq \alpha_{k,k}$.

Substituting Eqs.~\ref{eq:relation-greens-function-fourier-eg-ge}, \ref{eq:poralizability-tensor-static-limit}, and \ref{eq:inxs-elements-of-polarizability-tensor} into the sum of the response functions in Eqs.~(\ref{eq:response-functions-R3}) and (\ref{eq:response-functions-R4}) yields,
\begin{equation}\label{eq:sum-response-functions-R3-R4}
\begin{split}
    \tilde{R}_3 + \tilde{R}_4 &= - \sum_{e} \mathcal{G}_{eg}(\omega_{\rm opt}) \Big\{ \mathrm{Tr}_{\rm nuc}\Big[\hat{j}_{ge}(0)\hat{j}_{eg}(0) \hat{\Sigma}_g(\mathbf{q}, t_2) \rho_g\Big] \\
    &\qquad \qquad \qquad \qquad \qquad  - \mathrm{Tr}_{\rm nuc}\Big[\hat{\Sigma}_g(\mathbf{q}, t_2) \hat{j}_{ge}(0) \hat{j}_{eg}(0) \rho_g\Big] \Big\} \\
    &= -\frac{i}{2} \Big(\mathrm{Tr}_{\rm nuc}\big[ \rho_g\hat{\tilde{\alpha}}(0) \hat{\Sigma}_g(\mathbf{q}, t_2) \big] - \mathrm{Tr}_{\rm nuc}\big[ \rho_g \hat{\Sigma}_g(\mathbf{q}, t_2) \hat{\tilde{\alpha}}(0)  \big]\Big)\\
    &= \mathrm{Im}\ \mathrm{Tr}_{\rm nuc}\Big[ \rho_g \hat{\alpha}(0) \hat{\Sigma}_g(\mathbf{q}, t_2) \Big],
\end{split}
\end{equation}
which allows us to rewrite the expression for the INXS signal as
\begin{equation}\label{eq:trxd-signal-qm-expt-approx}
\begin{split}
    \Delta S(\mathbf{q}, \tau) &= \mathcal{N}^{\rm qm}  \mathrm{Re}  \int dt_2\  |E_{\rm opt}(\tau -t_2)|^2 \big[  \tilde{R}_4(\mathbf{q}, t_2) +  R_3(\mathbf{q}, \mathbf{t})  \big]\\
    &= \mathcal{N}^{\rm qm}  \mathrm{Im}  \int dt_2\   |E_{\rm opt}(\tau -t_2)|^2   \mathrm{Tr}_{\rm nuc}\Big[ \rho_g \hat{\alpha}(0) \hat{\Sigma}_g(\mathbf{q}, t_2) \Big],
\end{split}
\end{equation}
where one may take the expression for $\hat{\Sigma}(\mathbf{q})$ in Eq.~\ref{eq:scattering-operator-IAM} that we outlined in SI Sec.~\ref{ssec:xray-scattering} or combine the treatment outlined in this subsection with a different treatment of $\Sigma(\mathbf{q})$. 

Equation \ref{eq:trxd-signal-qm-expt-approx} is a fully quantum mechanical expression for the time-resolved scattering experiment with optically induced anisotropy. To arrive at this expression, we have combined a linear response approach with approximations inspired by our experimental setup. This includes assuming the impulsive nature of the X-ray interaction with matter, thermal equilibrium of the system prior to the light-matter interactions, the fact that both X-ray and optical pulses are off-resonance with electronic excitations, and that the optical pulse interacts with the system via a Raman interaction. In addition, in Secs.~\ref{ssec:xray-scattering} and \ref{ssec:Raman-interaction}, we have outlined additional approximations inspired both by the INXS experiment and widely adopted in Raman spectroscopy and X-ray scattering, respectively. 

\subsection{Limit of classical nuclei}
\label{ssec:classical-nuclei-limit}

Since X-ray scattering and Raman spectroscopies can often be captured assuming that the nuclei evolve classically (i.e., are guided by Hamilton's equations), we now turn to the classical nuclear limit of the INXS signal, Eq.~\ref{eq:trxd-signal-qm-expt-approx}, we have derived above. The classical nuclei approximation has the additional advantage that it allows one to exploit the thoroughly developed molecular dynamics toolbox to simulate the INXS signal for complex systems. To arrive at the classical nuclei limit, we recall the expression for the sum of the $\tilde{R}_3$ and $\tilde{R}_4$ response functions in Eq.~\ref{eq:sum-response-functions-R3-R4} and note that this combination can be rewritten as the imaginary component of the equilibrium time correlation function of the polarizability, $\hat{\tilde{\alpha}}(0)$ and the scattering operator, $\hat{\Sigma}(\mathbf{q}, t_2)$
\begin{equation}\label{eq:sum-response-functions-R3-R4}
\begin{split}
    \tilde{R}_3 + \tilde{R}_4 &= \mathrm{Im}\ \mathrm{Tr}_{\rm nuc}\Big[ \rho_g \hat{\alpha}(0) \hat{\Sigma}_g(\mathbf{q}, t_2) \Big]\\
    &\equiv  C^{\prime \prime}_{\tilde{\alpha}\Sigma }(t_2).
\end{split}
\end{equation}

While one can take multiple paths to arrive at the classical limit of a quantum time correlation function \cite{Mukamel1995,Egorov1999,Craig2005,Ramirez2004}, we adopt the approach where one approximates the quantum mechanical Kubo-transformed equilibrium time correlation function by its classical analogue \cite{Craig2005,Ramirez2004},
\begin{equation}
\begin{split}
    C^{\rm Kubo}_{AB}(t)  &= \frac{1}{Z\beta}\int_0^{\beta} d\lambda\ \mathrm{Tr}\big(e^{-(\beta - \lambda) \hat{H}}\hat{A}e^{-\lambda \hat{H}} \hat{B}(t) \big)\\
    &\approx C^{\rm clas}_{AB}(t),
\end{split}
\end{equation}
where $\hat{B}(t) = e^{i\hat{H} t/\hbar}\hat{B}e^{i\hat{H} t/\hbar}$ the quantum and classical equilibrium time correlation functions take the forms
\begin{subequations}
\begin{align}
    C_{AB}(t) &= Z^{-1} \mathrm{Tr}\big( e^{-\beta \hat{H}} \hat{A}(0) \hat{B}(t)\big),\\
    C^{\rm clas}_{AB}(t) &= Z_{\rm clas}^{-1} \int \int d\mathbf{R} d\mathbf{P}\ e^{-\beta H(\mathbf{R}, \mathbf{P})} A(\mathbf{R}_0, \mathbf{P}_0) B(\mathbf{R}_t, \mathbf{P}_t),
\end{align}
\end{subequations}
and $Z = \mathrm{Tr}[e^{-\beta \hat{H}}]$ is the partition function of the system, $\mathbf{R}$ and $\mathbf{P}$ are the coordinates and momenta of the atoms in the system, $\mathbf{R}_0, \mathbf{P}_0$ are the positions and momenta at $t=0$ and $\mathbf{R}_t, \mathbf{P}_t$ are the positions and momenta at $t$ as evolved using Hamilton's equations of motion. This approach is motivated by the fact that the quantum mechanical Kubo-transformed equilibrium time correlation function and its classical analogue are both real and even functions of time and is an exact replacement for correlation functions of the position (of linear order) in harmonic systems \cite{Craig2004}. 

At this point, we exploit the fact that the Kubo-transformed time correlation function can be easily connected to the full quantum time correlation function and its imaginary and real components separately in frequency space. In particular, one can show that \cite{Berne1970,Ramirez2004,Craig2005}
\begin{subequations}
\begin{align}
    C''(\omega) &= \frac{\beta \hbar \omega}{2i} C^{\rm Kubo}(\omega). \label{eq:fourier-imag-part-tcf-vs-full-tcf}
\end{align}
\end{subequations}
where the frequency space version of the time correlation function is given by its Fourier transform,
\begin{equation}
    C(\omega) = \int_{-\infty}^{\infty} dt\ e^{-i\omega t} C(t).
\end{equation}
Transforming back to real space, we obtain the following expression for the imaginary component of a correlation function, $C''(t)$,
\begin{equation}\label{eq:imaginary-to-classical-tcf}
\begin{split}
    C''(t) &= \frac{\beta \hbar}{2 i } \int_{-\infty}^{\infty} d\omega\ e^{i\omega t}  \omega C^{\rm Kubo}(\omega)\\
    &= \frac{\beta \hbar}{2 i^2 } \frac{d}{dt} \int_{-\infty}^{\infty} d\omega\ e^{i\omega t} C^{\rm Kubo}(\omega)\\
    &= - \frac{\beta \hbar}{2} \frac{d}{dt} C^{\rm Kubo}(t)\\
    &\approx - \frac{\beta \hbar}{2 } \frac{d}{dt} C^{\rm clas}(t).
\end{split}
\end{equation}
We are now in a good position to employ Eq.~\ref{eq:imaginary-to-classical-tcf} to obtain the classical nuclei approximation to the sum of the response functions in Eq.~\ref{eq:sum-response-functions-R3-R4}
\begin{equation}\label{eq:sum-response-functions-R3-R4-classical}
\begin{split}
    \tilde{R}_3 + \tilde{R}_4 &= C^{\prime \prime}_{\tilde{\alpha}\Sigma }(t_2)\\
    &\approx - \frac{\beta \hbar}{2} Z_{\rm clas}^{-1} \int \int d\mathbf{R} d\mathbf{P}\ e^{-\beta H(\mathbf{R}, \mathbf{P})} \tilde{\alpha}(\mathbf{R}_0) \Sigma(\mathbf{q}, t_2).
\end{split}
\end{equation}
Substituting Eq.~\ref{eq:sum-response-functions-R3-R4-classical} into our quantum mechanical expression for the INXS signal, Eq.~\ref{eq:trxd-signal-qm-expt-approx}, yields an expression for the INXS signal in the limit of classical nuclei, which is directly compatible with the toolbox of (force field, machine-learned potential, or ab initio) molecular simulation,
\begin{equation}\label{eq:trxd-signal-qm-expt-approx-classical-nuclei}
\begin{split}
    \Delta S^{\rm clas}(\mathbf{q}, \tau) &= \mathcal{N}^{\rm clas} \int dt_2\   |E_{\rm opt}(\tau -t_2)|^2   C^{\rm clas}_{\tilde{\alpha}\Sigma}(t_2),
\end{split}
\end{equation}
where
\begin{equation}
    C^{\rm clas}_{\tilde{\alpha}\Sigma}(t_2) \equiv  \int \int d\mathbf{R} d\mathbf{P}\ \frac{e^{-\beta H(\mathbf{R}, \mathbf{P})}}{Z_{\rm clas}} \tilde{\alpha}(\mathbf{R}_{0})\Sigma(\mathbf{q}, t_2),
\end{equation}
and the proportionality factor is
\begin{equation}\label{eq:final-prop-factor-inxs-signal-classical-nuclei}
\begin{split}
    \mathcal{N}^{\rm clas} &= -\beta \mathcal{N}^{\rm qm}\\
    &= -\frac{\beta (\mathbf{e}_{\rm xr} \cdot \mathbf{e}_{\rm s})^2|E_{\rm s}|^2 |E_{\rm xr}|^2}{\omega_{\rm opt}^2 \omega_{\rm xr}^2 \omega_{\rm s}^2}.
\end{split}
\end{equation}
We note that in our proportionality factor, we have not retained numerical constants, e.g., $\hbar$ or factors of 2, but rather only kept the functional dependence of this prefactor on experimental parameters that can be used to determine whether our theoretical approach recovers the correct scaling with respect to the angle between the polarization of the incoming vs.~scattered X-rays, the frequencies of the different photons in the scattering experiment,  the power of the incoming and scattered X-ray fields, and the inverse thermal energy $\beta = [k_BT]^{-1}$. 

\section{Experimental details}
\label{app:exp_details}
\subsection{Experimental setup}
The INXS experiment was conducted at the XPP endstation of the Linac Coherent Lights Source (LCLS). The liquid chloroform sample (CHCl$_3$) at room temperature was excited using a linearly polarized 800 nm laser pulse and probed using the 8.8kev monochromatic X-rays pulses using the large offset diamond crystal monochromator (LODCM) available at XPP. The optical laser pulses had a duration of ~50 fs full-width at half-maximum (FWHM) and the X-ray pulses had a ~40 fs duration FWHM. Combined with the group velocity mismatch in the 50 $\mu$m sample jet the resulting time-resolution was measured to be ~70 fs using the fast anisotropic response from a liquid water sample. The 100ml chloroform sample was recirculated using a HPLC pump connected to a 50um diameter fused silica nozzle resulting in a 50 $\mu$m cylindrical chloroform jet ensuring replenished sample between each consecutive shot at 120Hz. The optical laser polarization was offset by a 30\textdegree angle with respect to the X-ray polarization in order to more easily distinguish the anisotropic INXS signal from the x-ray polarization in the raw scattering signal. Following each pulse a scattering image was recorded on the downstream CSPAD detector for further analysis.

\begin{figure}[h!]
    \begin{center}
        \includegraphics[width=0.5\textwidth]{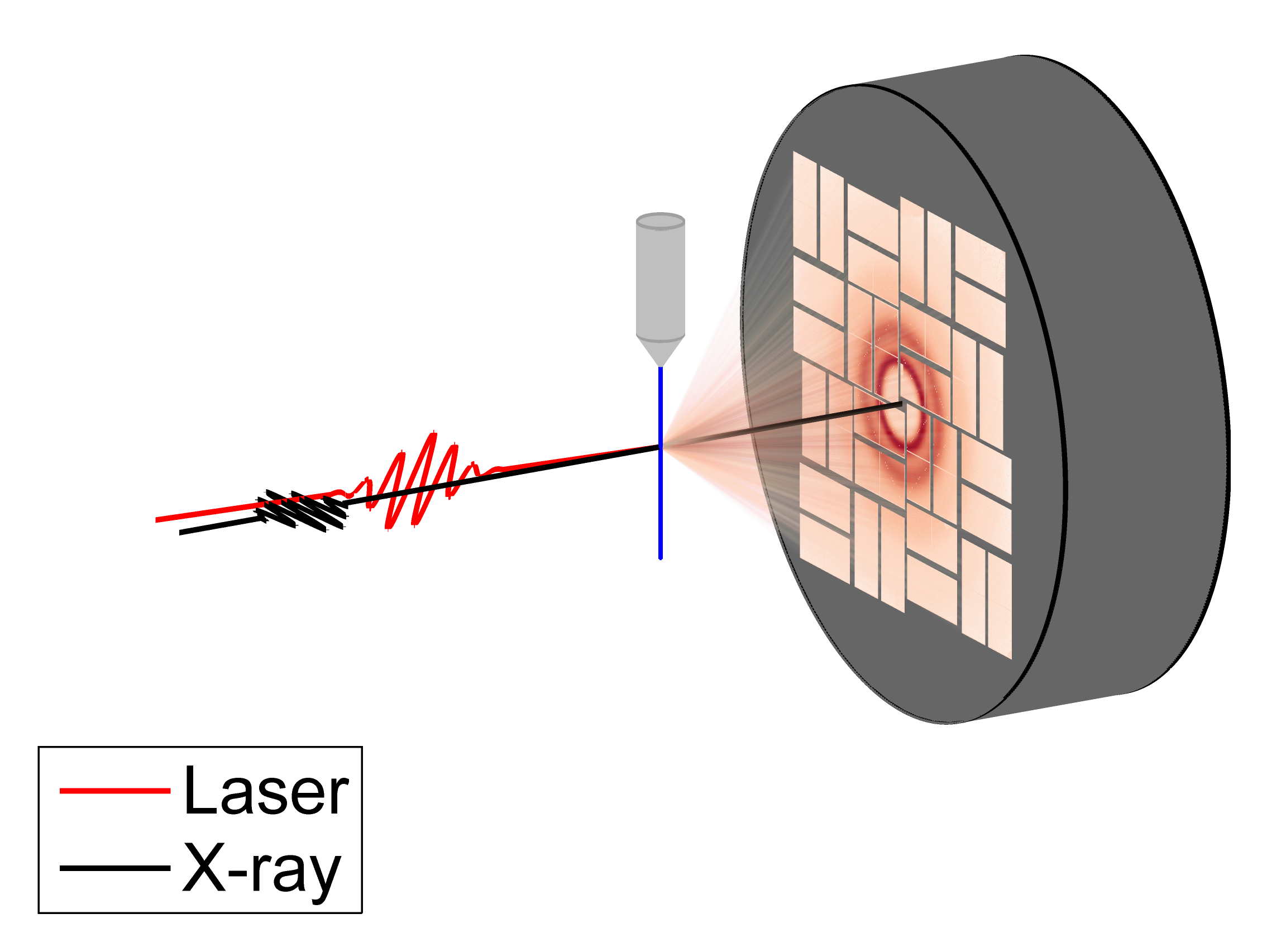}
    \end{center}  
    \caption{Schematic of the experimetal setup. The Nozzle producing the liquid Chloroform jet ~5cm from the CSPAD detector repeatedly pumped by 800nm laser pusles and probed by 8.8keV x-ray pulses at varying delays.}
	\label{fig:exp_schematic}
\end{figure}

\subsection{Data analysis}\label{ssec:exp-data-analysis}
The CSPAD detector images were recorded for each X-ray pulse. The detector images were filtered based x-ray intensity and first binned with respect to time delay (in 5~fs temporal bins) and then masked. The difference scattering ($\Delta S $) was subsequently constructed by subtracting the unpumped scattering data from $<$-200~fs and re-binned with respect to q, azimuthal angle ($\phi$) and delay. 21 $\phi$ bins, 80 time bins (25~fs/bin), and 55 q bins (0.84~\AA\textsuperscript{-1}/bin) were used.  Following the procedure in~\cite{Biasin2018}, $\Delta S_0$ and $\Delta S_2$ were extracted from a fit of the data to a linear combination of the zeroth and second order Legendre polynomials:
\begin{align}\label{eq:expsignal-s0-s2}
    \Delta S (q,t,\phi) = \Delta S_0(q,t) P_0(\cos(\phi))+\Delta S_2(q,t) P_2(\cos(\phi))
\end{align}

\subsection{Experimental sample heating}
The Isotropic signal that grows in on the ps timescale in $\Delta S_0$ contains the scattering signal resulting from a temperature increase of the solvent. On these ultra-fast timescales the signal can be described by the $\delta S/dT |\rho$ \cite{Kjaer2013}. Fitting the $\Delta S_0$ from 1-1.5ps identifies a temperature increase of 18.22K. This Increase in temperature originates from multiphoton excitation of the solvent. The high laser power was chosen to improve the amplitude of the anisotropic signal as experimental beamtime was limited. It is noted that the thermal signal is purely isotropic and typically grows in on the ps timescale depending on the solvent and therefore does not affect the signal in $\Delta S_2$.
\begin{figure}[h!]
    \begin{center}
        \includegraphics[width=0.85\textwidth]{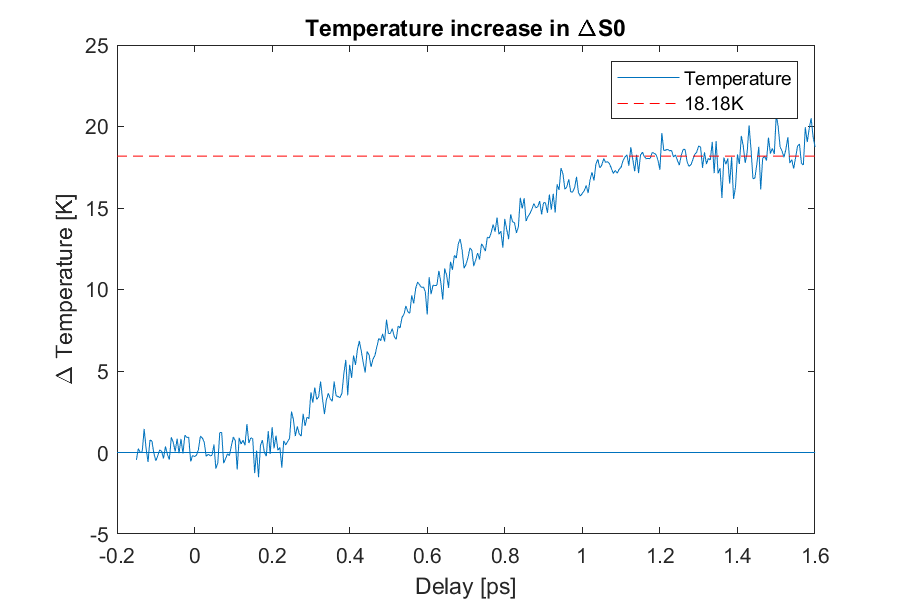}
    \end{center}  
    \caption{Temperature increase in $\Delta S_0$ as fit by the $\delta S/dT |\rho$ solvent heating signal from Ref.~\cite{Kjaer2013}}
	\label{fig:ExpS8}
\end{figure}

\subsection{Higher order anisotropy contributions to the experimental INXS signal}
In order to ensure higher order, nonlinear processes did not contribute to the INXS signal, we fit the data to a linear combination of Legendre polynomials up to eighth order:
\begin{align}
    \Delta S (q,t,\phi) = \sum_{n=0,2,4,6,8}  \Delta S_n(q,t) P_n(\cos(\phi))
\end{align}

The $\Delta S_0$ and $\Delta S_2$ components agree well with the second order fit, and the higher order $\Delta S_n$ components do not have any distinct features on the fast time scale corresponding to INXS, as seen in the principal components of each as a function of time (SI Fig.~\ref{fig:ExpS8}).Principal components were generated from a singular value decomposition of the data.  Based on this time dependence, we ascribe the higher order components to variations in the detected heating signal due to variations in the CSPAD detector behavior. These small fluctuations in the isotropic heating signal can appear as anisotropy, but their slow time evolution matches that of the isotropic signal.  Furthermore, any higher order anisotropy due to the optical pulse should decay on the order of picoseconds due to rotational dephasing. 

\begin{figure}[]
    \begin{center}
        \includegraphics[width=0.85\textwidth]{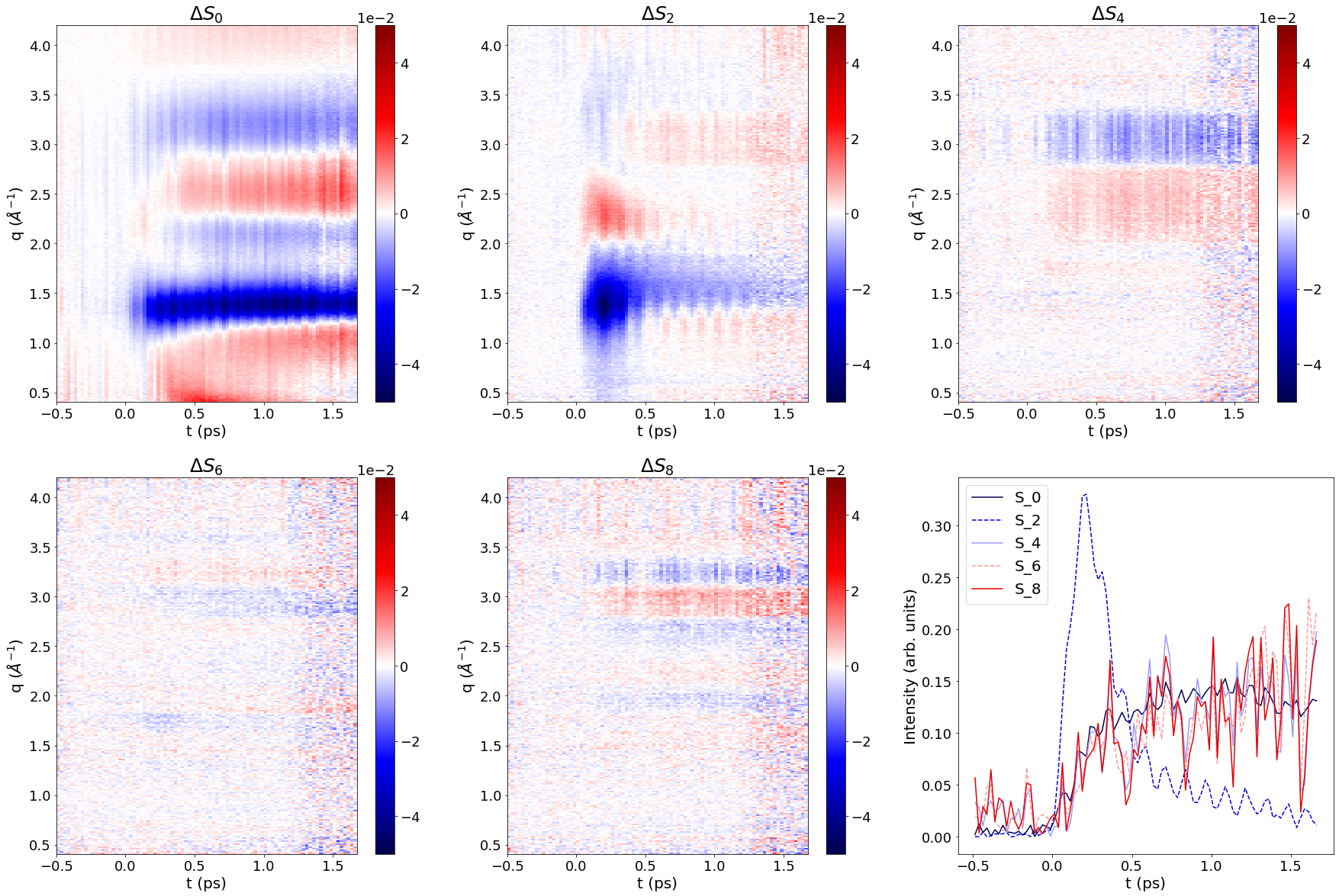}
    \end{center}  
    \caption{Decomposition of experimental difference scattering data using Legendre polynomials up to eighth order.  Also shown (bottom right) are the principal components of each $\Delta S_n$ as a function of time.}
	\label{fig:ExpS8}
\end{figure}

\section{Higher order anisotropy contributions to the simulated INXS signal}
\begin{figure}[h]
    \begin{center}
        \includegraphics[scale=0.85]{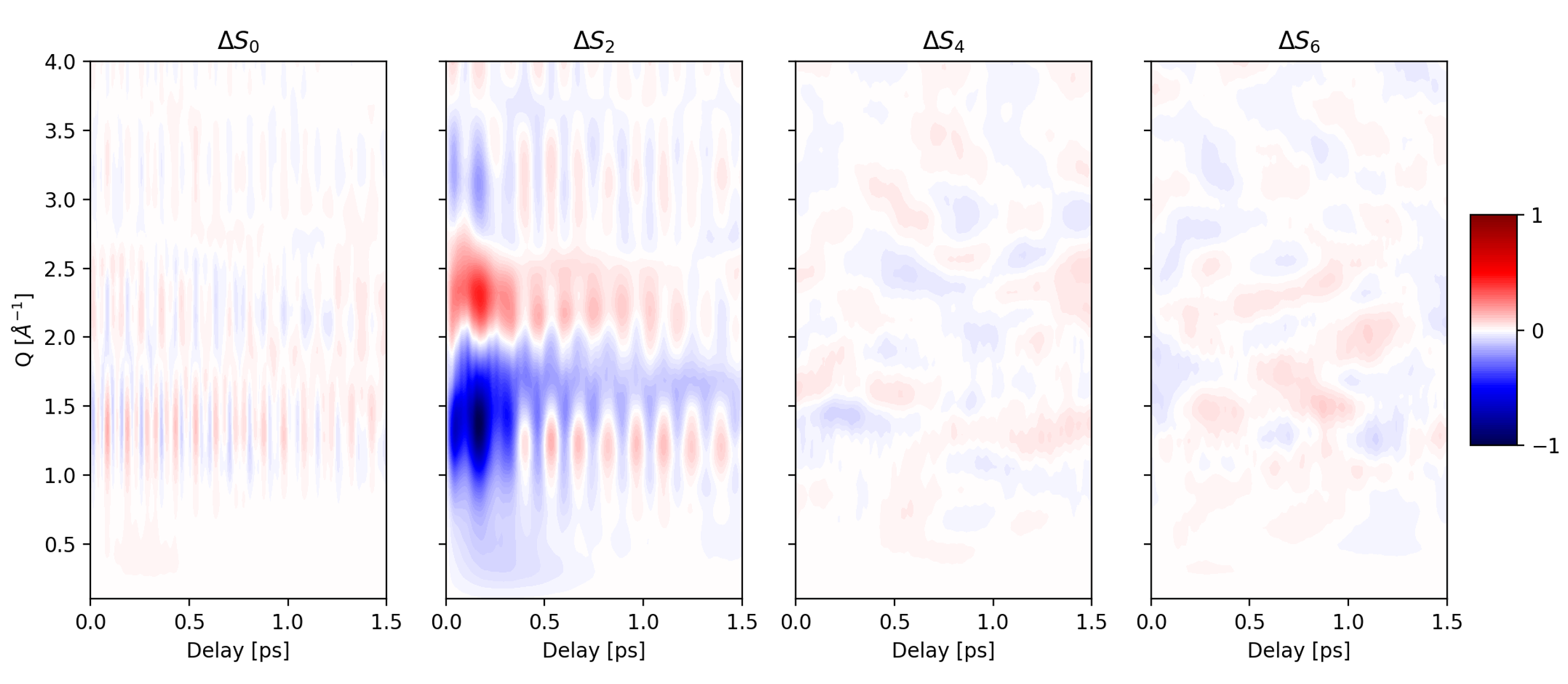}
    \end{center}
    \caption{Simulated $\Delta S_{0}(q,t)$, $\Delta S_{2}(q,t)$, $\Delta S_{4}(q,t)$, and $\Delta S_{6}(q,t)$ showing negligible contributions from the higher order anisotropic components of the INXS signal. Note that for these calculations, a 1~fs FWHM optical pulse was convolved with the INXS response.}
	\label{fig:s4_s6_sim_qt_1fs_all}
\end{figure}

\section{Simulation details}
\subsection{Molecular dynamics details}
The molecular dynamics (MD) simulations of liquid chloroform were performed using Tinker~\cite{Ponder2004} version 6.3.3 using an AMOEBA force field that was specifically parameterized for organochlorine compounds~\cite{Ponder2010,Mu2014}. The Packmol program~\cite{Martinez2009} was used to setup a periodically replicated cubic simulation cell with side lengths of 30.07~\AA~consisting of 203 chloroform molecules, which corresponds to a density of 1.48~g/cm$^3$ and is consistent with the experimental density of liquid chloroform at 298.15~K~\cite{Atik2007}. A total of 9 separate simulations were equilibrated for 50~ps in the NVT ensemble at 298 K, and then subsequently run for 1~ns using a 0.5~fs timestep. The Bussi-Parinello thermostat~\cite{Bussi2007} was employed, using a time constant of 0.1~ps for the initial equilibration runs and 1~ps for the following production runs. The smooth particle mesh Ewald (PME) algorithm~\cite{Essmann1995} using a real space cutoff of 9~\AA~was used for treating long-range electrostatic interactions. A separate cutoff of 12~\AA~was used for all other non-bonded potential energy interactions. At each timestep for these simulations, the atomic positions are written out and a custom version of Tinker's POLARIZE program was used to calculate the polarizability tensor for the simulation cell by applying an external electric field along each axis.

\subsection{Calculation of INXS time correlation function}
The calculation of the simulated INXS signal requires the construction of time correlation function specified in Eq.~2 of the main text (i.e. $\langle \alpha_{\boldsymbol{\epsilon}_{\rm opt}}(0, \{ \mathbf{R}\}) \Sigma(\mathbf{q}_{xy}, t) \rangle_{eq}$). We calculated $\Sigma(\mathbf{q}_{xy}=\left[q_x, q_y, 0\right], t)$ for a grid of values where $q_x$, and $q_y$ each spanned the domain $\left[-8.149~\text{\AA}^{-1},8.149~\text{\AA}^{-1}\right]$ with a spacing of $2\pi/30.07$~\AA$^{-1}$. The time correlation function was calculated out to a delay time of 7.5~ps with a resolution of 0.5~fs using all 9 separate 1~ns molecular dynamics trajectories we ran for liquid chloroform.

The isotropic nature of the material being sampled via our INXS experiment implies that it should be possible to cyclically rotate the coordinate system (i.e., $x \rightarrow y, y \rightarrow z, z \rightarrow x$ and $x \rightarrow z, y \rightarrow x, z \rightarrow y$), leading to equivalent results on the scattering planes ($x-y \rightarrow y-z$ and $x-y \rightarrow z-x$, respectively). We therefore exploit this invariance to 90$\circ$ rotations to improve our statistics, allowing us to calculate the time correlation function for when the optical pulse is linearly polarized in the $z-x$ and $y-z$ planes. This involves calculating $\Sigma(\mathbf{q}_{zx}=\left[q_x, 0, q_z\right], t)$ and $\Sigma(\mathbf{q}_{yz}=\left[0, q_y, q_z\right], t)$, and correlating them with $\tilde{\alpha}_{\boldsymbol{\epsilon}_{\rm opt}} = \cos^2(\phi) \alpha_{xx} + \sin^2(\phi) \alpha_{zz} + \sin(\phi)\cos(\phi)(\alpha_{xz} + \alpha_{zx})$ and $\tilde{\alpha}_{\boldsymbol{\epsilon}_{\rm opt}} = \cos^2(\phi) \alpha_{yy} + \sin^2(\phi) \alpha_{zz} + \sin(\phi)\cos(\phi)(\alpha_{yz} + \alpha_{zy})$, respectively. We use the average time correlation function of these three separate instances to calculate the INXS signal.

Since we are correlating small fluctuations of large quantities, from a numerical perspective it is advantageous to calculate the correlation function of the fluctuations and then correct for the long-time average. In particular, we are looking to calculate $\langle \tilde{\alpha}_{\boldsymbol{\epsilon}_{\rm opt}}(0, \{ \mathbf{R}\}) \Sigma(\mathbf{q}_{xy}, t) \rangle_{eq}$. We use the fluctuations, $\delta\tilde{\alpha}_{\boldsymbol{\epsilon}_{\rm opt}}(t, \{ \mathbf{R}\}) = \tilde{\alpha}_{\boldsymbol{\epsilon}_{\rm opt}}(t, \{ \mathbf{R}\}) - \langle\tilde{\alpha}_{\boldsymbol{\epsilon}_{\rm opt}}(\{ \mathbf{R}\})\rangle$ and $\delta\Sigma(\mathbf{q}_{xy}, t)= \Sigma(\mathbf{q}_{xy}, t)-\langle\Sigma(\mathbf{q}_{xy})\rangle$, to calculate
\begin{align}
    \langle \tilde{\alpha}_{\boldsymbol{\epsilon}_{\rm opt}}(0, \{ \mathbf{R}\}) \Sigma(\mathbf{q}_{xy}, t) \rangle_{eq} = \langle \delta\tilde{\alpha}_{\boldsymbol{\epsilon}_{\rm opt}}(0, \{ \mathbf{R}\}) \delta\Sigma(\mathbf{q}_{xy}, t) \rangle_{eq} + \langle\tilde{\alpha}_{\boldsymbol{\epsilon}_{\rm opt}}(\{ \mathbf{R}\})\rangle\langle\Sigma(\mathbf{q}_{xy})\rangle.
\end{align}

\subsection{Raman spectrum and normal mode analysis for the simulation}
\label{app:sim_raman}
\begin{figure}[h!]
    \begin{center}
        \includegraphics[scale=0.35]{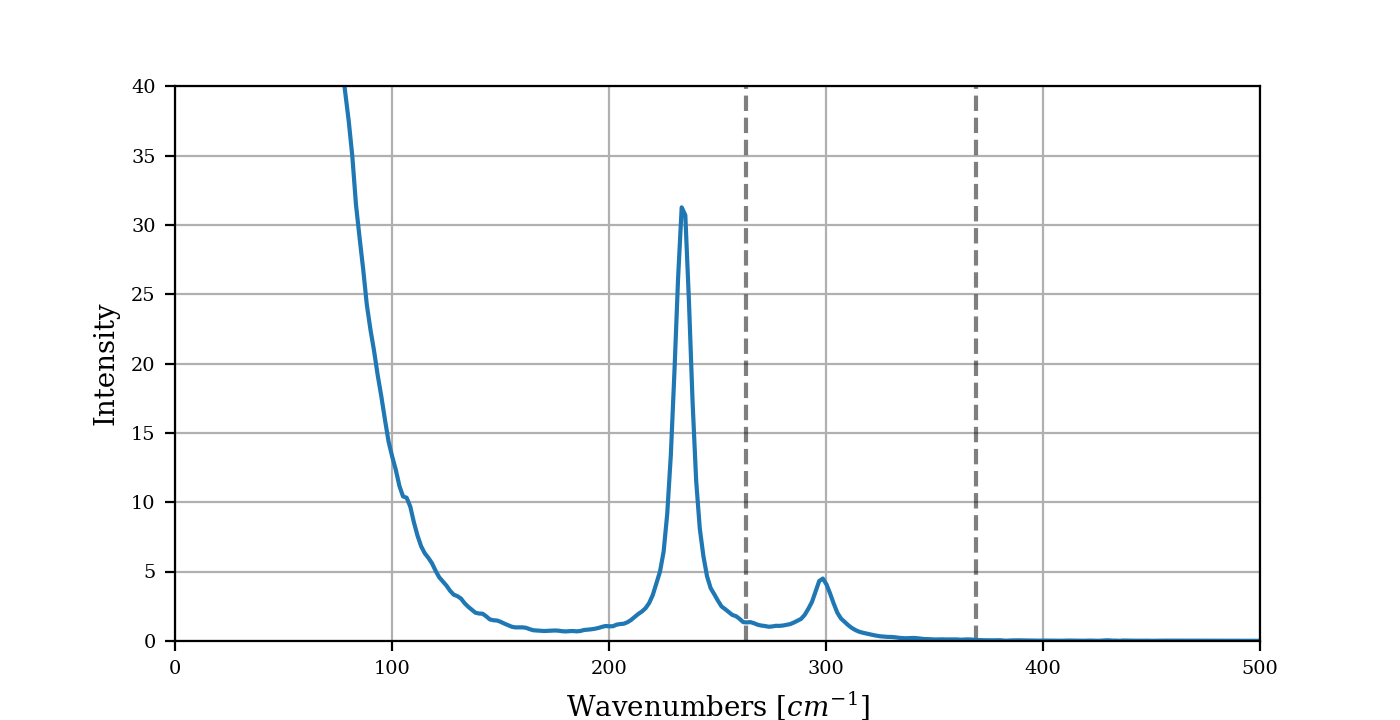}
    \end{center}
    \caption{Two lowest frequency peaks of the anisotropic Raman spectrum from our polarizable force field simulation of liquid chloroform (234 and 298 cm$^{-1}$). The two vertical dashed lines indicate the positions of the two lowest frequency experimental peaks (261 and 366 cm$^{-1}$)~\cite{Madigan1951} }
	\label{fig:sim_raman}
\end{figure} 

\begin{table}[h!]
    \centering
    \begin{tabular}{c|c|c|c|}
        Symmetry labels & NMA frequency (cm$^{-1}$) & Experimentally assigned frequency (cm$^{-1}$) \\ \hline\hline
        E  & 229  & 261  \\ \hline
        A1 & 293  & 363  \\ \hline
        A1 & 675  & 680  \\ \hline
        E  & 949  & 774  \\ \hline
        E  & 1219 & 1220 \\ \hline
        A1 & 2936 & 3034 \\ \hline
    \end{tabular}
    \caption{Vibrational frequencies from normal mode analysis (NMA) of a chloroform molecule using the AMOEBA force field~\cite{Mu2014} with assigned symmetry labels and comparisons with experimental gas-phase measurements and assignments from IR/Raman spectroscopy experiments~\cite{Shimanouchi1972,RudNielsen1942,Madigan1951,Gibian1951}}
    \label{tab:nma}
\end{table}

\section{Raw vs reconstructed experimental INXS signal}
\begin{figure}[h!]
    \begin{center}
        \includegraphics[width=0.85\textwidth]{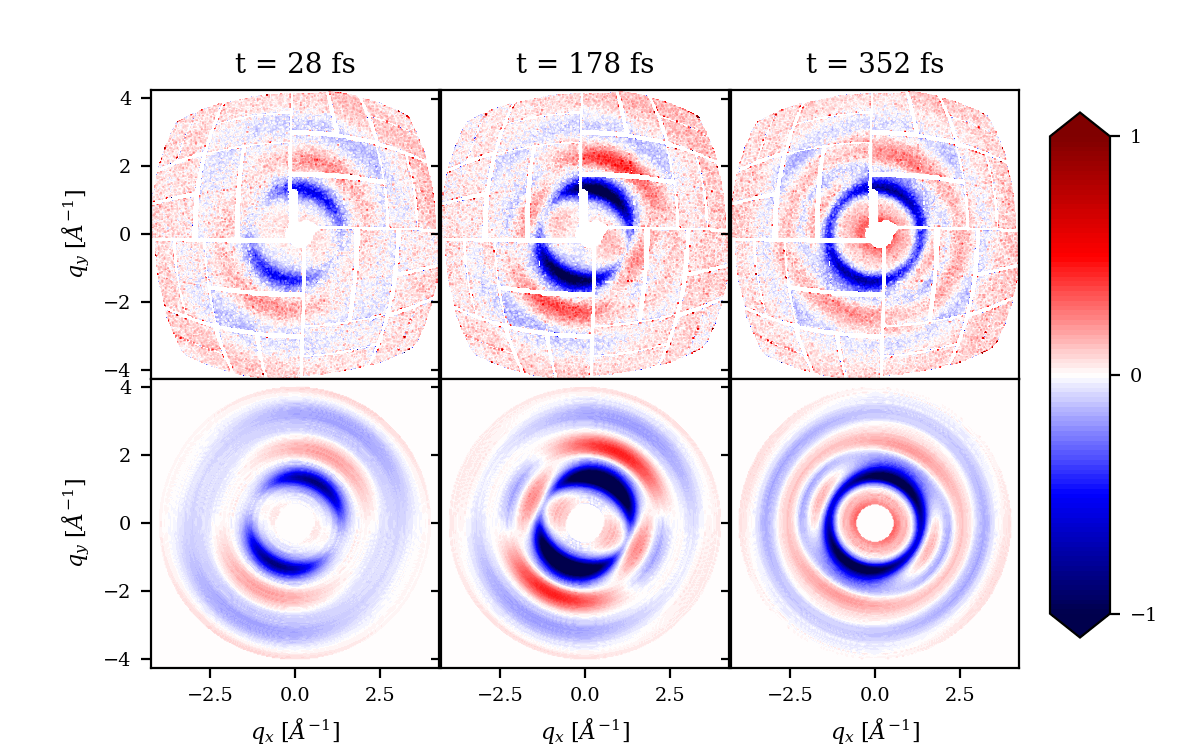}
    \end{center}  
    \caption{Comparison of the raw experimental signal (top row) vs. the reconstructed experimental signal (bottom row) for three different time delays. The reconstructed signal is calculated as the sum of the fitted $\Delta S_0(q,t)$ and $\Delta S_2(q,t)$ terms (see SI Sec.~\ref{ssec:exp-data-analysis}, SI Eq.~\ref{eq:expsignal-s0-s2}).}
	\label{fig:raw_vs_reconstruct}
\end{figure}

\section{Transformed S0 and S2 from time to frequency space}
In order to extract the frequency information from $\Delta S_0$ and $\Delta S_2$, singular value decomposition of the data is performed to isolate the high frequency component.  Any slowly varying features in the high frequency component are subtracted and this is then Fourier transformed into frequency space. 

\section{Decomposition of the INXS signal into isotropic and anisotropic components}
\label{app:s0_s2_decomp}

\subsection{Decomposing the signal in reciprocal space}
\begin{figure}[h]
    \begin{center}
        \includegraphics[scale=0.5]{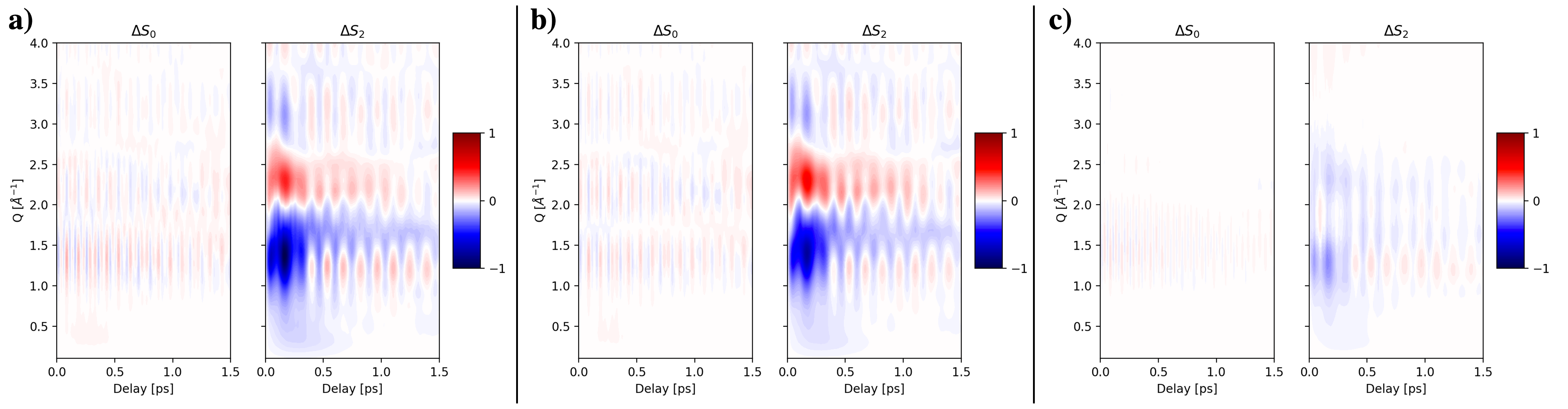}
    \end{center}
    \caption{Simulated $\Delta S_{0}(q,t)$ and $\Delta S_{2}(q,t)$ when accounting for (a) all pairwise distances, (b) only chlorine-chlorine pairwise distances, and (c) only non-chlorine pairwise distances in the calculation of the INXS signal. Note that for these calculations, a 1~fs FWHM optical pulse was convolved with the INXS response. Intensities are normalized using the same factor in all panels.}
	\label{fig:s0_s2_sim_qt_1fs}
\end{figure}

To asses the contributions of different symmetries to the total difference INXS signal we expand it in a Legendre polynomial basis, 
\begin{align}
    \Delta S(q, \phi, t) = \sum_{n=0}^{\infty}  \Delta S_{n}(q, t) P_{n}(\cos{\phi}),
\end{align}
where $P_n(x)$ is the $n$th Legendre polynomial. For the chloroform system studied here the contribution to the total signal from components with $n>2$ is small (see SI Fig.~\ref{fig:ExpS8} and SI Fig.~\ref{fig:s4_s6_sim_qt_1fs_all} for the experiment and simulation respectively) and hence the total difference signal is well approximated as,
\begin{align}
    \Delta S(q, \phi, t) \approx \Delta S_{0}(q, t) + P_{2}(\cos{\phi})\Delta S_{2}(q, t),
\end{align}
where we have used that $P_{0}(x)=1$.
The total signal for a system with atom types $a$ and $b$ can be separated based on different pairwise contributions as $\Delta S(q,\phi,t) = \sum_{ab}\Delta S_{ab}(q,\phi, t)$ and can be written when expanded in terms of Legendre polynomials as
\begin{align}
\label{eq:S_ab}
\sum_{ab}\frac{\Delta S_{ab}(q,\phi,t)}{F^*_a(q)F_b(q)} = \sum_{ab}\sum_{n=0}^{\infty} \frac{\Delta S_{ab,n}(q,t)}{F^*_a(q)F_b(q)}P_n(\cos\phi),
\end{align}
where $F_a(q)$ and $F_b(q)$ are the atomic form factors associated with elements $a$ and $b$. In the current experiment the form factors for chlorine are much larger than the other elements present (carbon and hydrogen) and hence as shown in SI Fig.~\ref{fig:s0_s2_sim_qt_1fs} the INXS signal is dominated by chlorine-chlorine scattering i.e. $a=b={\rm Cl}$ and Eq.~\ref{eq:S_ab} reduces to,
\begin{equation}
\label{eq:S_ab_Cl}
\frac{\Delta S(q,\phi,t)}{F^*_{Cl}(q)F_{Cl}(q)} = \sum_{n=0}^{\infty} \frac{\Delta S_n(q,t)}{F^*_{Cl}(q)F_{Cl}(q)} P_n(\cos\phi).
\end{equation}

\subsection{Transforming the decomposed signal to real space}
To express the difference signal in real space we can Fourier transform it with respect to $q$. The Fourier transform of the $n$th component of the signal is
\begin{equation}
\Delta g^{(n)}_{ClCl}(r,t) = \int^{\infty}_0 dq \: q^2 \int^{\pi}_0 d\varphi \sin\varphi \: e^{i\boldsymbol{q}\cdot\boldsymbol{r}} \frac{\Delta S_n(q,t)}{F^*_{Cl}(q)F_{Cl}(q)} P_n(\cos\varphi).
\end{equation}
Defining the angle between $q$ and $r$ to be $\varphi$ we get
\begin{equation}
\Delta g^{(n)}_{ClCl}(r,t) = \int^{\infty}_0 dq \: q^2 \int^{\pi}_0 d\varphi \sin\varphi \: e^{iqr\cos\varphi} \frac{\Delta S_n(q,t)}{F^*_{Cl}(q)F_{Cl}(q)} P_n(\cos\varphi).
\end{equation}
Note that for generality, we have chosen $\varphi$ which may or may not coincide with $\phi$, the angle for the polarization of the optical pulse in the $x$-$y$ plane.
The expression above can be further simplified by making the change of variable $x=\cos\varphi$ and using the identity
\begin{equation}
    j_n(y) = \frac{1}{2i^n}\int^{1}_{-1}dx \: e^{ixy}P_n(x),
\end{equation}
where $j_n$ is the $n$th spherical Bessel function, to get
\begin{equation}
\Delta g^{(n)}_{ClCl}(r,t) = 2 i^{n} \int^{\infty}_0 dq \: q^2 j_n(qr) \frac{\Delta S_n(q,t)}{F^*_{Cl}(q)F_{Cl}(q)}.
\end{equation}
As stated above, the signal is assumed to be into to a $n=0$ and $n=2$ component. Looking at the $n=0$ component we see that
\begin{equation}
\Delta g^{(0)}_{ClCl}(r,t) = 2 \int^{\infty}_0 dq \: q^2 \frac{\sin(qr)}{qr} \frac{\Delta S_0(q,t)}{F^*_{Cl}(q)F_{Cl}(q)},
\end{equation}
which is the difference radial distribution function. Likewise the $k=2$ component is given by
\begin{equation}
\Delta g^{(2)}_{ClCl}(r,t) = -2 \int^{\infty}_0 dq \: q^2 j_2(qr) \frac{\Delta S_2(q,t)}{F^*_{Cl}(q)F_{Cl}(q)},
\end{equation}
which can be viewed as a generalization of the difference radial distribution function that contains angularly averaged information of the signal with the second-order Legendre polynomial.

More generally, we can associate $g^{(n)}_{ClCl}(r,t)$ as the nth order Legendre polynomial projection of an angularly resolved pair distribution function $g_{ClCl}(r,\varphi, t)$, i.e.
\begin{equation}
     \Delta g^{(n)}_{ClCl}(r,t) = \int_{0}^{\pi} \text{d}\varphi~\sin{\varphi}\Delta g_{ClCl}(r,\varphi, t) P_{n}(\cos{\varphi}).
\end{equation}
\begin{equation}
    \Delta g_{ClCl}(r,\varphi,t) = \sum_{n}\Delta g^{(n)}_{ClCl}(r,t) P_n(\cos\varphi).
\end{equation}

\subsection{Performing the decomposition of the simulated signal}
The isotropic $\Delta S_0$ and anisotropic $\Delta S_2$ contributions of the experimental signal were be determined via a linear regression fit~\cite{Biasin2016,Biasin2018,Kjaer2019,Lorenz2010,Baskin2005,Baskin2006,Ki2021}. For the simulated signal, we instead numerically integrate a linearly interpolated version of the 2D simulated difference scattering signal,

\begin{align}
    \Delta S_{n}(q, t) &= \frac{2n+1}{2}\int_{0}^{\infty}\text{d}\phi\sin(\phi)P_{n}(\cos(\phi))\Delta S(q, \phi, t).
\end{align}

For high resolution and low noise simulated data, the two decomposition methods (the one here and the method detailed in SI Sec.~\ref{ssec:exp-data-analysis}) give similar results. However, the linear regression fit is better suited for experimental data that contains detector artifacts and other features.

\section{Decomposition of the simulated INXS signal into chlorine-only contribution}
\label{app:cl_only_signal}

\section{Intramolecular and intermolecular simulated $\Delta g_{ClCl}^{(0)}(r,t)$ and $\Delta g_{ClCl}^{(2)}(r,t)$}
\label{app:g0_g2_intra_inter}
\begin{figure}[h]
    \begin{center}
        \includegraphics[scale=0.5]{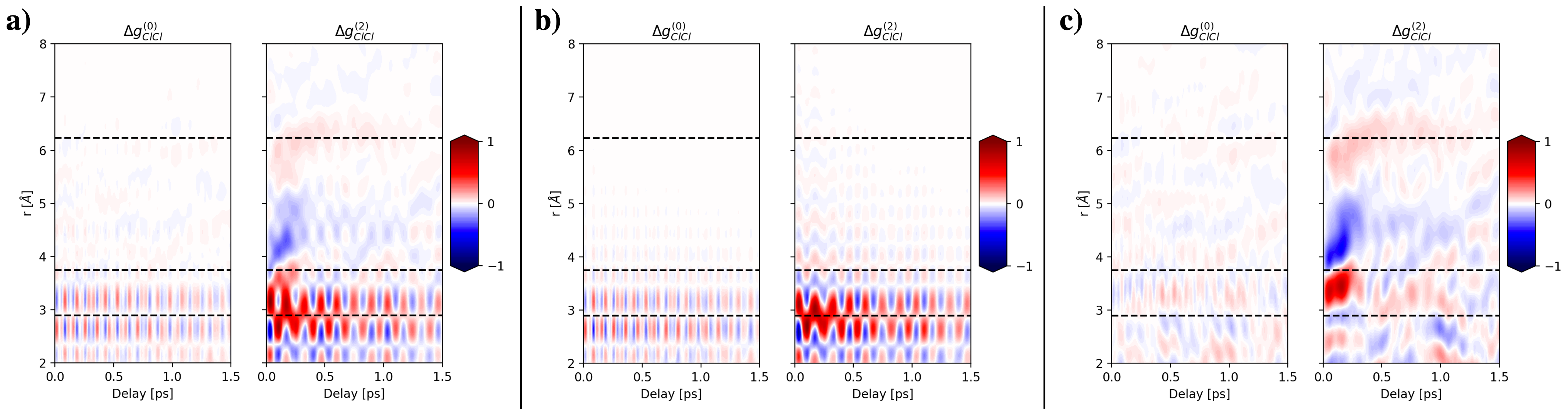}
    \end{center}
    \caption{Simulated $\Delta g_{ClCl}^{(0)}(r,t)$ and $\Delta g_{ClCl}^{(2)}(r,t)$ when accounting for (a) all pairwise chlorine-chlorine distances, (b) only intramolecular chlorine-chlorine pairwise distances, and (c) only intermolecular chlorine-chlorine pairwise distances. Note that for these calculations, a 1~fs FWHM optical pulse was convolved with the INXS response.}
	\label{fig:g0_g2_sim_rt_1fs_clcl}
\end{figure} 

\section{Angular dependence of the radial distribution function}

The radial distribution function provides knowledge about the density fluctuations as a function of the interparticle distance $r$. Here, our goal is to learn of the angular fluctuations that contribute the observed signal at a given interparticle distance $r$. INXS provides angularly resolved information due to the symmetry breaking induced by the directional Raman interaction it provides access to information that is not available from direct X-ray scattering experiments. To provide insight into these angular fluctuations we thus construct a quantity containing the angular deviation from the isotropic response (encoded in the radial distribution function, $g_{ab}^{(0)}(r)$), which by construction vanishes upon angular averaging,
\begin{equation}
    \label{eq:int_g_r_theta}
    \int^{\pi}_0 d\varphi\ \sin\varphi \tilde{g}_{ab}(r,\varphi)=0,
\end{equation}
where 
\begin{equation}
    \tilde{g}_{ab}(r,\varphi) = g_{ab}(r,\varphi) -\frac{1}{2}g_{ab}^{(0)}(r),
\end{equation}
with the factor of $1/2$ ensuring that the angular average is zero, and 
\begin{equation}
    g_{ab}(r,\varphi) = \frac{\langle\sum^{N_a}_{k=1}\sum^{N_b}_{l=1}\delta(r_{kl}-r)\delta(\varphi_{kl}-\varphi)\rangle}{4N_aN_b\pi r^2 \sin\varphi},
\end{equation}
is the angularly resolved radial distribution i.e. that is defined at a particular distance and angle.

While a direct X-ray scattering experiment would only give access to the isotropic response, where the angularly averaged $\tilde{g}_{ab}(r,\varphi)$ would be zero, the anisotropic part of the INXS signal provides access to higher order moments. As such the anisotropic INXS signal is expected to have the most pronounced features at distances $r$ where the variations of $g_{ab}(r,\varphi)$ as a function of $\varphi$ from its angularly averaged value, $g_{ab}^{(0)}(r)$, are large. This can be characterized using the variance of $g_{ab}(r,\varphi)$ from $g_{ab}^{(0)}(r)$ for a given distance $r$ defined as
\begin{align}
    \sigma^2_{\varphi,ab}(r) &= \int^{\pi}_0 d\varphi\ \sin\varphi \tilde{g}_{ab}^2(r,\varphi) - \left(\int^{\pi}_0 d\varphi\ \sin\varphi \tilde{g}_{ab}(r,\varphi)\right)^2 \\
    &= \int^{\pi}_0 d\varphi\ \sin\varphi \tilde{g}_{ab}^2(r,\varphi),
\end{align}
where the second lines follows from Eq.~\ref{eq:int_g_r_theta}.

\bibliography{bibliography}